\documentclass[twocolumn,trackchanges]{aastex63}
\usepackage[figure,figure*]{hypcap}
\usepackage{multirow}
\usepackage{natbib}
\usepackage{xcolor}

\hypersetup{linkcolor=red,citecolor=cyan,filecolor=magenta,urlcolor=blue}

\newcommand{\sex}{\textsc{SExtractor}}

\newcommand{\photutils}{\textsc{PhotUtils}}
\newcommand{\beagle}{\textsc{Beagle}}
\newcommand{\cloudy}{\textsc{Cloudy}}
\newcommand{\pyneb}{\textsc{PyNeb}}
\newcommand{\python}{\textsc{Python}}

\newcommand{\ha}{H$\alpha$}
\newcommand{\hb}{H$\beta$}
\newcommand{\pa}{Pa$\alpha$}

\newcommand{\oiia}{\textsc{[Oii]}$\lambda$3726}
\newcommand{\oiib}{\textsc{[Oii]}$\lambda$3729}
\newcommand{\oiiia}{\textsc{[Oiii]}$\lambda$4959}
\newcommand{\oiiib}{\textsc{[Oiii]}$\lambda$5007}
\newcommand{\oiiic}{\textsc{[Oiii]}$\lambda$4363}

\shorttitle{Clumpy galaxies in SDSS S82}
\shortauthors{Mehta et al.}

\begin{document}

\title{Investigating Clumpy Galaxies in the Sloan Digital Sky Survey Stripe 82 using the Galaxy Zoo}

\author[0000-0001-7166-6035]{Vihang Mehta}
\affiliation{Minnesota Institute for Astrophysics, University of Minnesota, 116 Church Street SE, Minneapolis, MN 55455, USA}
\email{mehta074@umn.edu}

\author{Claudia Scarlata}
\affiliation{Minnesota Institute for Astrophysics, University of Minnesota, 116 Church Street SE, Minneapolis, MN 55455, USA}

\author{Lucy Fortson}
\affiliation{Minnesota Institute for Astrophysics, University of Minnesota, 116 Church Street SE, Minneapolis, MN 55455, USA}

\author{Hugh Dickinson}
\affiliation{School of Physical Sciences, The Open University, Milton Keynes, MK7 6AA, UK}

\author{Dominic Adams}
\affiliation{Minnesota Institute for Astrophysics, University of Minnesota, 116 Church Street SE, Minneapolis, MN 55455, USA}

\author{Jacopo Chevallard}
\affiliation{Sorbonne Universit{\'e}, UPMC-CNRS, UMR7095, Institut d'Astrophysique de Paris, F-75014, Paris, France}

\author{St{\'e}phane Charlot}
\affiliation{Sorbonne Universit{\'e}, UPMC-CNRS, UMR7095, Institut d'Astrophysique de Paris, F-75014, Paris, France}

\author{Melanie Beck}
\affiliation{Minnesota Institute for Astrophysics, University of Minnesota, 116 Church Street SE, Minneapolis, MN 55455, USA}

\author{Sandor Kruk}
\affiliation{European Space Agency (ESA), European Space Research and Technology Centre (ESTEC), Keplerlaan 1, 2201 AZ Noordwijk, The Netherlands}

\author{Brooke Simmons}
\affiliation{Physics Department, Lancaster University, Lancaster, LA1 4YB, UK}

\begin{abstract}
Giant, star-forming clumps are a common feature prevalent amongst high-redshift star-forming galaxies and play a critical role in shaping their chaotic morphologies and yet, their nature and role in galaxy evolution remains to be fully understood. A majority of the effort to study clumps has been focused at high redshifts, and local clump studies have often suffered from small sample sizes. In this work, we present an analysis of clump properties in the local universe, and for the first time, performed with a statistically significant sample. With the help of the citizen science-powered \textit{Galaxy Zoo: Hubble} project, we select a sample of 92 $z<0.06$ clumpy galaxies in Sloan Digital Sky Survey Stripe 82 galaxies. Within this sample, we identify 543 clumps using a contrast-based image analysis algorithm and perform photometry as well as estimate their stellar population properties. The overall properties of our $z<0.06$ clump sample are comparable to the high-redshift clumps. However, contrary to the high-redshift studies, we find no evidence of a gradient in clump ages or masses as a function of their galactocentric distances. Our results challenge the inward migration scenario for clump evolution for the local universe, potentially suggesting a larger contribution of \textit{ex-situ} clumps and/or longer clump migration timescales.
\end{abstract}


\section{Introduction}

The study of galaxy morphology and visual inspection of astrophysical objects have proven invaluable for furthering our understanding of galaxy formation and evolution. While the nearby and low-redshift galaxies are mostly dominated by \textit{spiral} and \textit{elliptical} morphologies, observations of the high-redshift universe over the past few decades have revealed a different picture of galaxy demographics. Typical star-forming galaxies (SFGs) in the low-redshift universe have a smooth, disk-like appearance, while their high-redshift counterparts exhibit much more chaotic and irregular morphologies \citep[e.g.,][]{driver95,driver98,glazebrook95,vandenbergh96,conselice05,elmegreen05b,elmegreen07,elmegreen09,scarlata07,guo12,guo15,guo18,shibuya16,soto17,fisher17}.

A standout feature amongst high-redshift SFGs is the presence of bright knots or ``clumps.'' These clumps appear as bright concentrations of light that make up the irregular morphologies of the high-redshift galaxies. Physically, the clumps are likely sites of intense star-formation within the galaxy and hence are most evident in rest-frame UV images. Since their initial discovery, significant effort has been invested in studying these clumps in galaxies over a wide redshift range using rest-frame UV \citep[e.g.,][]{guo12,guo15,guo18,livermore12,shibuya16,soto17,zavadsky18,messa19}, rest-frame optical \citep[e.g.,][]{conselice04,elmegreen09,schreiber11} as well as using emission lines such as \ha\ \citep[e.g.,][]{fisher17,zanella19,whitmore20}, and \pa\ \citep[e.g.,][]{larson20}, and CO (e.g., \citealt{swinbank10,liu20}; but also see \citealt{ivison20}).

The prevalence of clumps and \textit{clumpy} galaxies has been shown to increase toward higher redshifts with a peak at around $z\sim2$ \citep{guo15,shibuya16}. Clumps studied so far have stellar masses that span $10^{7-9} M_\odot$ \citep[e.g.,][]{elmegreen07,guo12,guo18,zanella19} with specific star-formation rates (sSFR) that are elevated by a few orders of magnitude above the typical integrated value for their host galaxies \citep[e.g.,][]{genzel08,genzel11,guo12,guo18,hemmati14,fisher17}. The physical size of clumps is still a topic of debate in the literature. Unlensed surveys of high-redshift clumps have estimated the clump sizes to be $\sim$1 kpc \citep[e.g.,][]{elmegreen07,schreiber11}. However, even at Hubble Space Telescope (HST) resolution, the high-redshift clumps are likely to be unresolved. Clump studies done using lensed galaxies have found sizes as small as $\sim$30-100 pc \citep[e.g.,][]{livermore12,adamo13,cava18,zick20} even in high-redshift galaxies. It has been suggested that imaging resolution and sensitivity can introduce systematic biases in the inferred estimates of clumps sizes and masses \citep{fisher17,tamburello17,meng19}.

These star-forming clumps clearly play a critical role in shaping the morphologies of galaxies, most likely affect their evolution, and possibly are also tied to the mass assembly of galaxies. While the physical processes responsible for the formation of these clumps remains to be fully determined, two primary avenues of forming clumps have been proposed. Gravitational instabilities in gas-rich galactic disks caused due to turbulence can lead to formation of clumps \citep[e.g.,][]{bournaud07,bournaud09,dekel09,ceverino10,ceverino12,dekel14,inoue16}. The onset of this type of violent disk instability (VDI) can be due to the intense inflow of cold gas along narrow filaments \citep{birnboim03,keres05,dekel06,dekel09} potentially leading to fragmentation of gaseous disks and ultimately \textit{in-situ} formation of clumps. Clump formation driven by VDI is supported by observed properties of clumps in galaxies \citep[e.g.,][]{elmegreen07,bournaud08,guo15,guo18,fisher17}. On the other hand, clumps can also be formed \textit{ex-situ} due to galaxy interactions and mergers \citep[e.g.,][]{hopkins13,mandelker17}, which also finds support from some observational studies \citep[e.g.,][]{puech09,puech10,wuyts14,straughn15,ribeiro17,zanella19}. \cite{guo15} performed a large-scale clump study using a mass-complete galaxy sample over $0.5<z<3$ from HST CANDELS/GOODS-S and UDS fields and found evidence for VDI-dominated clump formation for massive galaxies ($\log{(M/M_\odot)}>9.8$), while they suggest merger-driven formation of clumps for galaxies with stellar masses of $9<\log{(M/M_\odot)}<9.8$.

Significant computational effort has also been invested into investigating the nature and behavior of clumps in cosmological simulations \citep[e.g.,][]{inoue14,inoue19,mandelker14,mandelker17,tamburello17,meng19} that have further posed the question of clump evolution. The ultimate fate of these clumps is also unknown. Two plausible clump evolution scenarios have been discussed in the literature: (i) long-lived clumps that end up migrating inward and eventually merge into the central bulge of the galaxy \citep{bournaud07,elmegreen08,mandelker14,mandelker17}, and (ii) short-lived clumps that are disrupted due to outflows from their own starburst or galactic shocks and/or outflows, and end up dissolving over a relatively short timescale ($\lesssim$50 Myr) and instead contribute to the thick disk \citep{bassett14,inoue14}. Some observational studies support the former idea of long-lived clumps based on the observed galacto-radial gradient in the colors of clumps \citep[e.g.,][]{schreiber11,guo12,guo18,shibuya16,soto17}, whereas other studies suggest the latter scenario of short-lived clumps that are quickly disrupted \citep[e.g.,][]{murray10,genel12,hopkins12,hopkins14,buck17,oklopcic17}. A follow-up clump study by \cite{guo18} also found a negative age gradient, consistent with the inward migration scenario, for their $0.5<z<1$ and $1<z<2$ samples.

Much of the observational effort invested in clump studies has been focused at redshifts $z\gtrsim0.5$. This is partly due to readily available high-resolution, rest-frame UV imaging for higher redshifts from HST, which is useful for studying clumps. However, the prevalence of clumpy galaxies is also higher in the early universe. As \citet{guo15} and \citet{shibuya16} report, the overall fraction of clumpy galaxies ($f_{clumpy}$) rises toward higher redshifts, reaching a peak of $\sim$60\%\ at $z\sim2$ before gradually declining. Toward lower redshifts, $f_{clumpy}$ drops much more sharply, reaching $\sim$40\%\ by $z\sim1$. Studies of clumps in the local universe have been limited to sample sizes of $\sim$10 galaxies \citep[e.g.,][]{fisher17,messa19}, since clumpy galaxies are inherently rare at low redshifts.

The citizen science-powered \textit{Galaxy Zoo} project \citep{lintott08,fortson12,willett13} has been widely successful at morphologically classifying all galaxies in the Sloan Digital Sky Survey (SDSS). The third phase of Galaxy Zoo featured the \textit{Galaxy Zoo: Hubble} (GZH) project \citep{willett17}, which primarily focused on classifying high-redshift HST galaxies, specifically including a set of questions asking the volunteers to classify the clumpy structure of the galaxies. In addition to the HST galaxies, it also included a sample of Stripe 82 galaxies from SDSS DR7 in order to provide a local sample for comparison. In this work, we use the subset of SDSS Stripe 82 galaxies combined with the volunteer classifications regarding the clumpy nature of the galaxies from GZH to select and study a sample of clumpy galaxies in the local universe ($z<0.06$). Specifically, we aim to compare the properties of clumps in the local universe with those in high-redshift ($z>0.5$) studies. We also investigate the gradients in clump masses and ages to distinguish between the clump formation and evolution scenarios.

This manuscript is arranged as follows: Section~\ref{sec:data} describes the data and selection of our clumpy galaxy sample used for this analysis; Section~\ref{sec:clump_finder} outlines our automated clump finder algorithm; Section~\ref{sec:measured_props} describes the clump photometry  as well as the estimation of morphological parameters and stellar population properties; Section~\ref{sec:results} presents our results along with a discussion; Section~\ref{sec:conclusions} wraps up with some concluding remarks.

Throughout this analysis, we use the \citet{planck15} cosmology and all magnitudes are expressed in the AB system \citep{oke83}.

\section{Data}
\label{sec:data}

\subsection{Galaxy Zoo: Hubble}
\label{sec:data_gzh}

\begin{figure*}[!ht]
\centering
\includegraphics[width=0.95\textwidth]{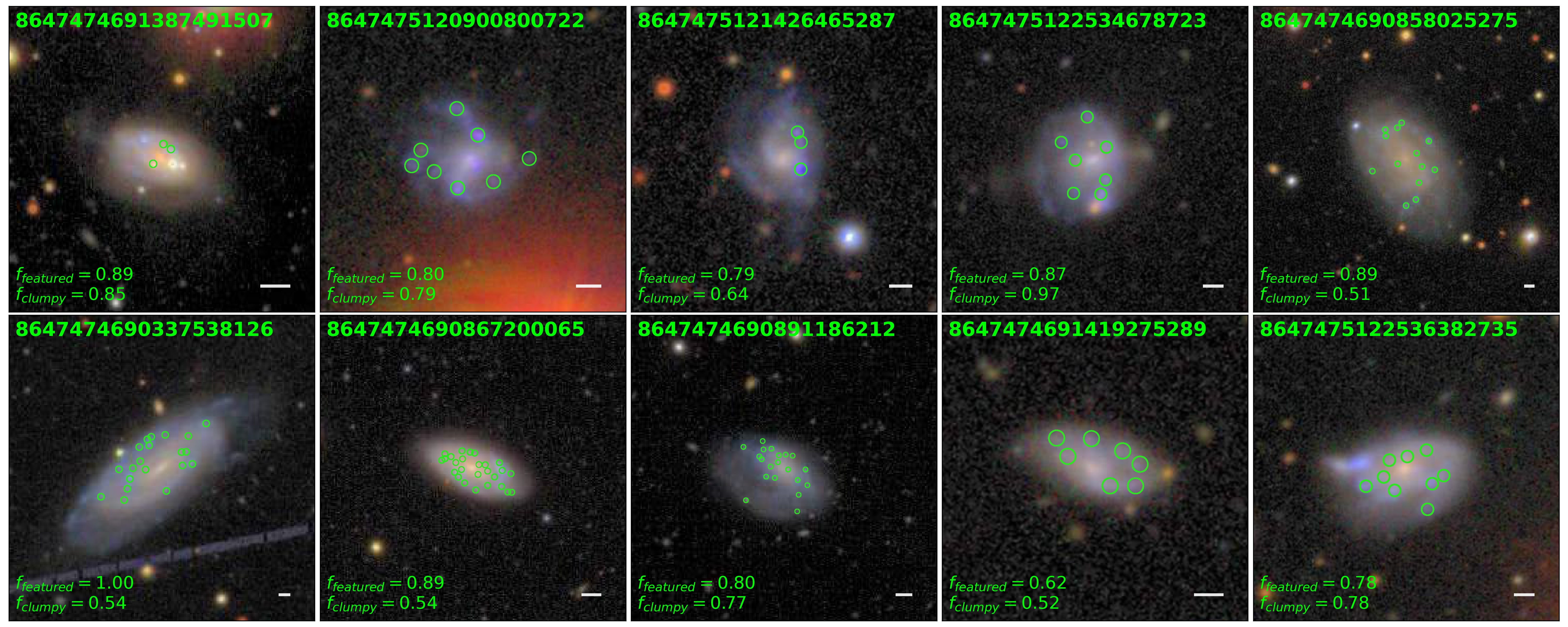}
\caption{A few examples from our sample of Stripe 82 clumpy galaxies selected from GZH. The RGB stamps shown here are made using the $gri$-band coadd-Stripe 82 images. The SDSS DR7 object identification number is denoted for each stamp as well as the featured and clumpy vote fractions from GZH. The white bar in the bottom right corner denotes 5\arcsec. The locations of all identified clumps are marked with \textit{green circles}.}
\label{fig:sample_examples}
\end{figure*}

In this work, we study a sample of local clumpy galaxies in SDSS selected with the help of the \textit{Galaxy Zoo: Hubble} project \citep{willett17}. While the primary goal of the \textit{Galaxy Zoo: Hubble} (GZH) project was to classify images from various \textit{Hubble} imaging surveys such as AEGIS, GOODS and COSMOS fields, it also included a sample of 21,522 single-epoch and 30,339 coadd-epoch images from SDSS DR7's coverage of Stripe 82. Since clumpy morphologies are known to be common at high redshifts, the GZH decision tree specifically included questions asking the volunteers to ascertain if the galaxy has a clumpy appearance provided it was already classified to be featured.

For this work, we apply cuts based on the volunteer responses to these questions to select local clumpy galaxies from the GZH Stripe 82 sample. Specifically, we consider two vote fractions:
\begin{itemize}
    \item Featured vote fraction, $f_{featured}$: The fraction of volunteers that voted a subject as having ``features or disk'' in response to the question ``Is the galaxy simply smooth and rounded, with no sign of a disk?''.
    \item Clumpy vote fraction, $f_{clumpy}$: The fraction of volunteers that voted affirmatively to the question ``Does the galaxy have a mostly clumpy appearance?''.
\end{itemize}
We consider a galaxy to be clumpy if the weighted $f_{featured}\geq0.5$ and weighted $f_{clumpy}\geq0.5$. Additionally,  we require that at least 20 volunteers have responded to the clumpy question in order to ensure that the clumpy vote fraction, $f_{clumpy}$, is statistically significant.

The volunteers separately classified galaxies identified in both single- as well as coadd-epoch images. If included in samples from both epochs, the same galaxy can have different vote fractions since volunteers classify them independently. Moreover, since the stamps are generated using photometric and morphological properties of the target galaxy, the single- and coadd-epoch stamps presented to the volunteers can be substantially different -- e.g., different measured Petrosian radii that are used to set the stamp size \citep{willett13,willett17} or a galaxy being deblended into multiple photometric objects, which are used as centers for the stamps. As a result, the Stripe 82 single- and coadd-epoch clumpy vote fractions in GZH can be different for the same galaxy. In order to be as inclusive as possible, we include all galaxies that pass the above-mentioned cuts in samples from either epoch. We take additional care to de-duplicate cases where multiple photometric objects are identified within the same galaxy. Overall, these cuts result in a sample of 453 unique clumpy galaxies spanning redshifts up to $z=0.2$.

\subsection{Additional cuts}
\label{sec:resolution_cuts}

Given that the physical sizes of known clumps range between $\sim40-100$ pc as observed in high-resolution lensed surveys \citep{livermore12,livermore15,fisher17,messa19}, the clumps observed in the GZH-selected sample are likely unresolved at the SDSS imaging resolution. However, in order to draw comparisons with other large-area surveys of clumps at higher redshift \citep[e.g.,][]{guo15,guo18}, we need to restrict our sample to where the physical scales probed are comparable. The \citet{guo15,guo18} studies are based on HST imaging and cover a redshift range over $1.5<z<3$, probing down to $\sim$1 kpc. Consequently, we choose to restrict our GZH-selected sample to $z\lesssim0.06$ where the SDSS point-spread function (PSF; $\sim$1\arcsec-1.2\arcsec) resolves physical scales $\lesssim$1 kpc, resulting in a sample that is comparable to the \citet{guo15,guo18} high-redshift clump studies. This further reduces our sample down to 125 galaxies.

\begin{figure}[!h]
\centering
\includegraphics[width=0.48\textwidth]{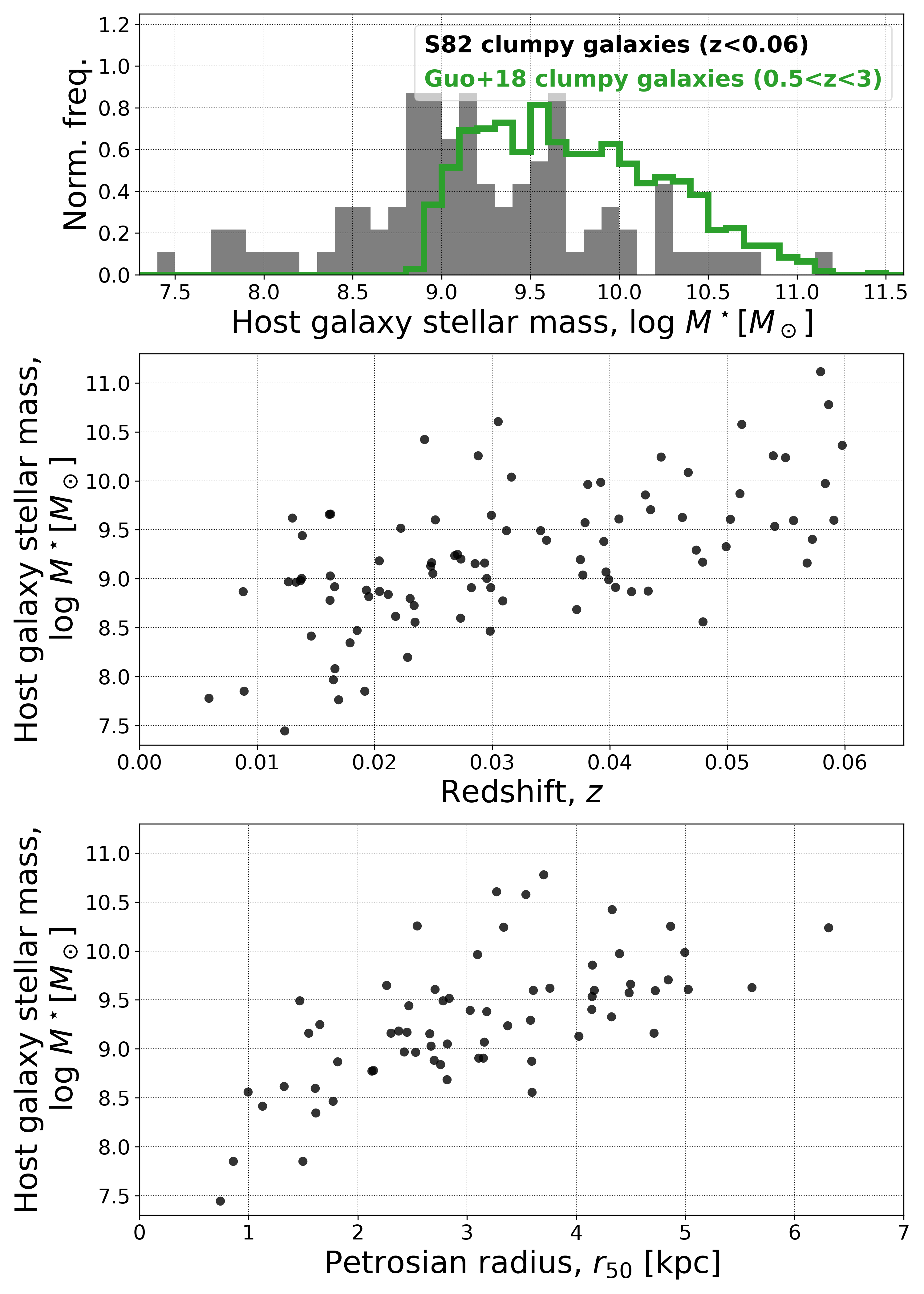}
\caption{\textit{Top panel:} The stellar mass distribution for the S82 clumpy galaxies shown in \textit{black} alongside the \citet{guo18} sample.  While we do not impose any cuts on the host galaxy flux or stellar mass in this work, the original GZH S82 sample was selected to have $r$-band apparent magnitude, $r<17.77$ \citep{willett17}. On the other hand, the \citet{guo18} sample imposed a $H$-band flux cut of $H_{160}<24.5$, which translates to a stellar mass cut of $\log{(M^\star/M_\odot)}\sim9$ at $z\sim2$. \textit{Middle panel:} The galaxy stellar mass as function of redshift for S82 clumpy sample. We specifically restrict the sample to $z<0.06$ to match the physical scales probed with the SDSS imaging resolution to that of the HST-based \citet{guo18} $0.5<z<3$ clump study (see Section~\ref{sec:resolution_cuts} for details). \textit{Bottom panel:} The galaxy stellar mass as a function of the SDSS-reported $r$-band Petrosian radius (in kpc) for the S82 clumpy sample.}
\label{fig:sample_props}
\end{figure}

From a visual inspection of this sample, it is obvious that there still remains a small fraction of contaminants such as galaxies in various merging states (15 cases), galaxies with multiple nuclei (2 cases), galaxies with overlapping foreground stars (13 cases), and even tight groups of elliptical galaxies in some cases (3 cases) that are misclassified as clumpy galaxies. Hence, a round of visual vetting by experts is warranted and after removing the contaminants, we are ultimately left with a sample of 92 $z<0.06$ clumpy galaxies selected from SDSS Stripe 82 with the help of GZH volunteers. Figure~\ref{fig:sample_examples} shows a few examples of clumpy galaxies from our sample and Figure~\ref{fig:sample_props} shows their stellar mass and redshift distributions.

\subsection{Imaging and spectroscopic data}

While the clumpy galaxy sample considered in this work includes galaxies identified as clumpy in either the single- or coadd-epoch images (see Section~\ref{sec:data_gzh}), coadd-epoch imaging is available for all targets. For all analysis presented in this work, we use the coadd-epoch Stripe 82 images to utilize the deepest available data for our targets. Specifically, we use the sky-subtracted, calibrated images (``corrected'' frames) for Stripe 82 coadd runs 106 and 206 available\footnote{\url{http://das.sdss.org/www/html/imaging/dr-75.html}} as part of SDSS DR7. Furthermore, we query the SDSS DR14 metadata to obtain the latest redshift measurements for our sample. One of the goals of this work is to investigate the gas-phase metallicities of clumps (Section~\ref{sec:clump_metallicity}). For this purpose, we search and retrieve all available spectra from SDSS DR14 within 1.5\arcmin\ of our target galaxies so that they can be position matched with any clumps that we find within the target galaxy.

\section{Clump Identification}
\label{sec:clump_finder}

\subsection{Automated clump finder}

\begin{figure}[!ht]
\centering
\includegraphics[width=0.48\textwidth]{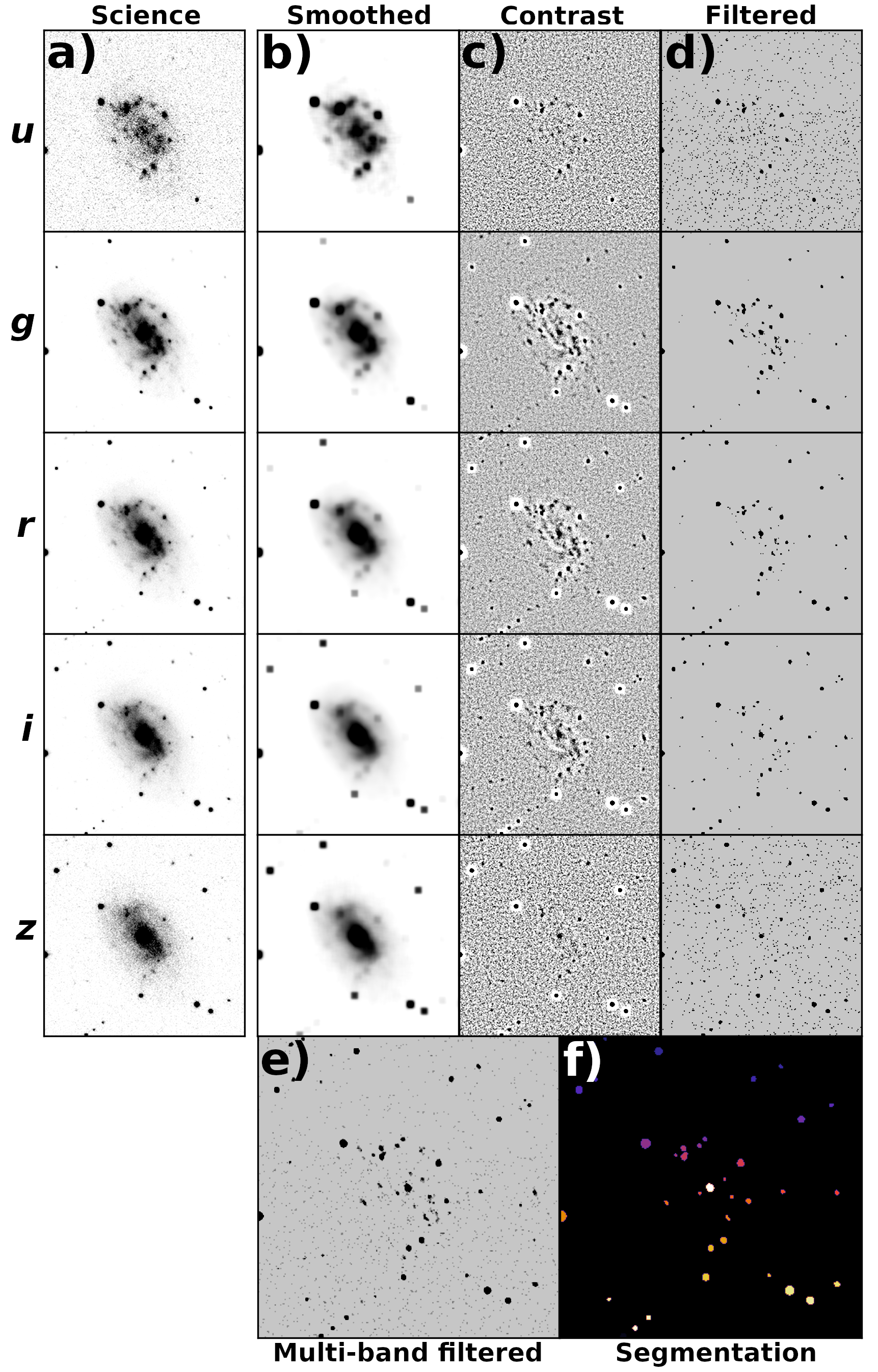}
\caption{Illustration of the clump finding algorithm. The figure shows the $ugriz$ science images for SDSS objID\#8647474690858025275 in panel $a$ ($ugriz$ arranged from top to bottom). The boxcar-smoothed, contrast and filtered images are shown in panels $b, c, d$, respectively. Panel $e$ shows the combined multi-band filtered image. Lastly, the segmentation map created by running \sex\ on the multi-band filtered image is shown in panel $f$, which shows all clump-like objects identified within the stamp.}
\label{fig:clump_finder}
\end{figure}

We identify the clumps in our clumpy galaxy sample using an automated approach. We adopt the contrast detection technique from \cite{guo15} with some modifications. Figure~\ref{fig:clump_finder} gives an overview of how the clump detection algorithm functions. We start by cutting out a stamp for each galaxy in the $ugriz$ filters from the SDSS ``corrected'' frames. For the sizes of the stamps, we use the angular size equivalent to 75 kpc with a minimum stamp size of 30\arcsec$\times$30\arcsec and a maximum stamp size of 120\arcsec$\times$120\arcsec. We opt against using the measures of galaxy sizes (e.g., Petrosian radii) because the complex morphologies of the galaxies in the initial sample render these measures unreliable in some cases. The automated clump detection then proceeds as follows.

First, we smooth the stamps for each galaxy with a boxcar filter to create a smoothed image. The size of the kernel used for this step has significant impact on the size of the clumps that are detected. The goal of the smoothed image is to smooth the point sources such that they can be identified in the contrast image. Hence, the kernel size has to be large enough to smooth the instrument PSF sufficiently. On the other hand, too large of a kernel size will identify larger substructures and also diminishes the algorithm's ability to separate individual clumps. We experiment with the kernel size to optimize the clump detection and find the kernel size of 10 pixels (3.96\arcsec) to be ideal for identifying the unresolved clumps in SDSS-resolution images.

Next, we generate a contrast image by subtracting the smoothed image from the original science image. We then filter the contrast image to mask out all pixels below 2$\sigma$ to create a filtered image. The $\sigma$ calculation is done after a 3$\sigma$ clipping to prevent image artifacts or nearby bright sources from affecting the filtering process. The final clump detection can then be performed by running a source detection algorithm on the filtered image.

However, the detection of clumps is strongly dependent on the filters used. The star-forming clumps stand out much more significantly in the $u$-band compared to the $griz$-bands. On the other hand, the $u$-band is much shallower compared to the $gri$-bands. Hence, we opt for a ``multi-band clump detection'' approach. The steps outlined above to create smoothed, contrast and filtered images are followed for each $ugriz$-band stamps. We then sum combine the $ugri$ filtered images to create a multi-band filtered image, which combines the individual $ugri$-band filtered images by performing a straight summation across the filters. This approach preserves the prominence of the clumps from the $u$-band while exploiting the added depth from the $gri$-bands. We exclude the $z$-band here since it does not add any significant benefit when identifying the blue star-forming clumps. On the other hand, it increases the risk of including foreground stars and red background galaxies, in addition to the fact that it is comparatively shallow. Finally, we run \sex\ on the multi-band filtered image to identify clump-like objects, requiring an absolute detection threshold, a minimum detection area of 5 pixels and a strong deblending threshold.

\subsection{Determining clump membership}
\label{sec:clump_membership}

\begin{figure*}[!ht]
\centering
\includegraphics[width=0.7\textwidth]{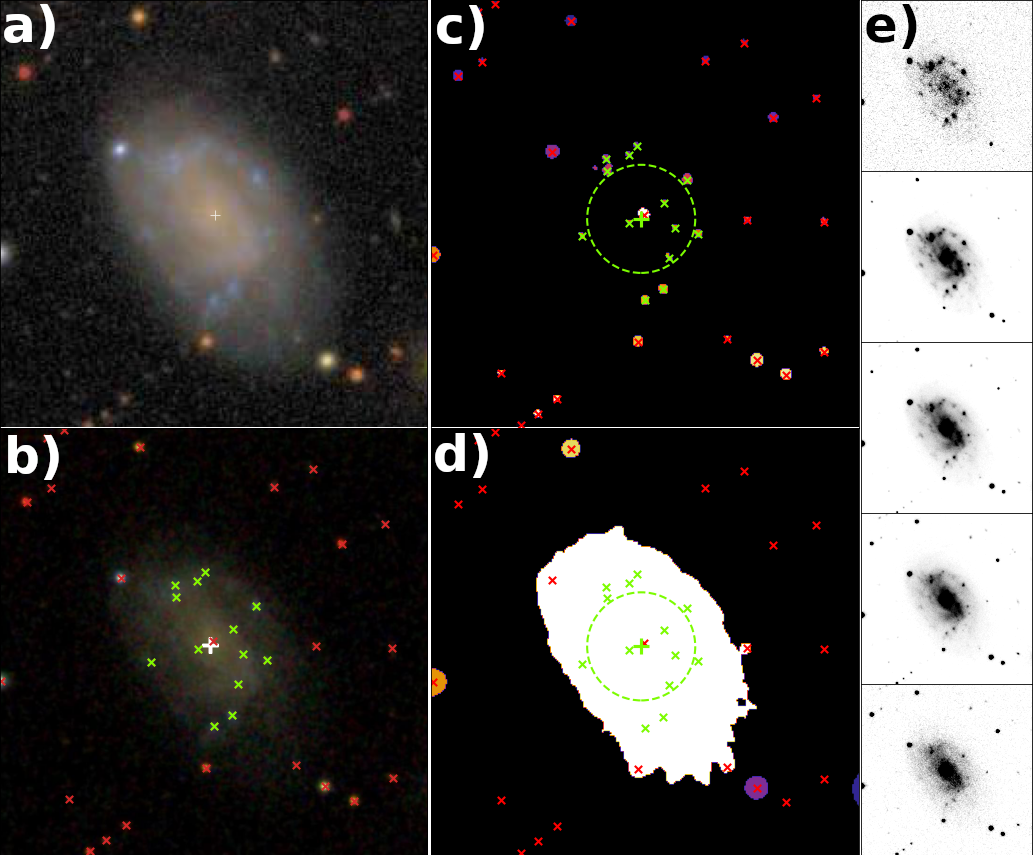}
\caption{Illustration of how clump membership is determined as well as the impact of visual vetting of the auto-detected clumps. Panel $a$ shows the Stripe 82 coadd-epoch RGB stamp (made using $gri$-bands) for SDSS objID\#8647474690858025275 (same object as in Figure~\ref{fig:clump_finder}); whereas Panel $b$ shows the DR14 (single-epoch) RGB stamp along with the clump locations overplotted. Panel $c$ shows the segmentation map generated from the multi-band filtered image that highlights all detected clump-like objects within the science image. Panel $d$ shows the segmentation map generated from the smoothed science image to determine the extent of the host galaxy. Panel $e$ shows the individual $ugriz$ stamps (from top to bottom) for the same object. The clumps that are designated as being real and part of the host galaxy are marked with \textit{green crosses}, whereas the rejected sources are marked in \textit{red crosses}, which include both objects that lie outside the extent of the galaxy as well as contaminants that are misidentified as clumps.}
\label{fig:clump_membership}
\end{figure*}

The clump finder algorithm described above merely identifies the location of the sources that look clump-like, i.e., a concentrated blob of flux that stands out from its immediate surroundings. This algorithm identifies all clump-like sources in the image, regardless of whether they are associated with the primary galaxy. In order to draw any conclusions about the clump properties, it is essential that we first ensure that the clumps under consideration are in fact real and associated with the host galaxy.

We establish the membership of the identified clumps based on their location with respect to the host galaxy. To do so, we run \sex\ on the smoothed $r$-band stamp to generate a segmentation map that outlines the extent of the host galaxy. Only clumps that are identified within the segmentation of the host galaxy are considered for analysis. Moreover, some of the host galaxies have pronounced central bulges that are often mistakenly identified as clumps. We also exclude any clumps identified at the location of the galaxies' center of mass (CoM; see Section \ref{sec:gal_props}).

\subsection{Additional visual vetting}
Foreground stars as well as background galaxies are potential contaminants for the sample of identified clumps since they can appear as point-like sources similar to the real clumps. In order to remove these, the automatically detected clumps undergo an additional round of expert visual vetting. The $gri$ false-color images are reinspected with all identified clump locations overlaid and any stars or background galaxies identifiable in the false-color image are excluded from the final clump sample.

Figure~\ref{fig:clump_membership} illustrates the membership determination (Section~\ref{sec:clump_membership}) as well as visual vetting process for automatically identified clumps for an example galaxy. These additional steps ensure that our clump sample is as pure as possible. The final sample consists of 543 clumps identified in our sample of 92 $z<0.06$ Stripe 82 galaxies.

\section{Measured Clump Properties}
\label{sec:measured_props}

\subsection{Clump Photometry}
\label{sec:photometry}
For all clumps identified within a target host galaxy, its $ugriz$-band photometry is measured from the SDSS \textit{corrected} (sky-subtracted and calibrated) frames. Here, we assume that the clumps are unresolved sources, which is reasonable given that with SDSS's $\sim$1\arcsec\ seeing, the imaging resolution reaches $\sim$1 kpc physical sizes at the redshifts of our target sample.

Under this assumption, the photometry is simplified as we can directly use a fixed aperture size for the clumps. However, the seeing can still vary from target to target for the SDSS imaging. Hence, we adopt an aperture size of 1.5 $\times$ the seeing FWHM corresponding to the imaging data quality for the specific host galaxy. We use the average of the $ugriz$ seeing FWHM as reported in the SDSS photometry catalog to set the aperture size. The aperture size is kept constant across the $ugriz$-bands for all clumps within a given host galaxy. The median seeing for the Stripe 82 coadd-epoch imaging for our sample is 1.3\arcsec, which spans over a range of 1.2\arcsec$-$1.4\arcsec.

We specifically use a small aperture size in order to minimize flux contamination from the host galaxy as well as from other clumps or astrophysical objects that may lie near the clump. Given that we assume the clumps to be unresolved, we can correct the measured fluxes for the flux loss due to the small aperture sizes. Assuming a Gaussian profile for the PSFs, we multiply the fluxes by a factor of $\sim$1.26 for the aperture correction. All photometry is done after masking all other identified clumps within the target galaxy.

The measured fluxes are also treated for galactic extinction. We correct the fluxes using the \citet{schlegel98} dust maps\footnote{Specifically, we use the Python implementation available at \url{https://github.com/adrn/SFD} to query the SFD dust maps.}. The reddening $E(B-V)$ is queried at each clump location and converted to an extinction (A$_\lambda$) for the $ugriz$ filters assuming a \citet{cardelli89} extinction law.

Since the clumps lie within their host galaxies, their fluxes are contaminated by the underlying galaxy light. It is critical to subtract this diffuse component from the clump photometry to recover accurate clump fluxes and to avoid misinterpretation of the results \citep[e.g.,][]{buck17,oklopcic17}. In order to quantify the diffuse galaxy light contribution, we measure the flux within an annulus of 2$-$3 $\times$ the seeing FWHM centered at the clump location and compute the median background level within. This background flux is then subtracted off from the clump photometry aperture.

The efficiency of the diffuse galaxy light subtraction can vary among individual clumps. There are several factors that complicate this process: the morphology of the galaxy may not be constant or smoothly varying on the scale of the background annulus size; some clumps may be extended and/or resolved; and neighboring clumps may have different flux ratios or distances. All of these can impact the diffuse light subtraction process. However, it is still critical to remove this component in order to ensure that the inferred physical properties of the clumps (e.g., mass, star-formation rate) are accurate. We verify the accuracy of the clump photometry including the diffuse light subtraction using mock clumps inserted into SDSS galaxies and ensure that the recovered clump photometry is unbiased and accurate down to the $\sim$0.1 mag. These clump simulations are fully described in an upcoming paper (Adams et al., in preparation).

\begin{figure*}[!ht]
\centering
\includegraphics[width=0.95\textwidth]{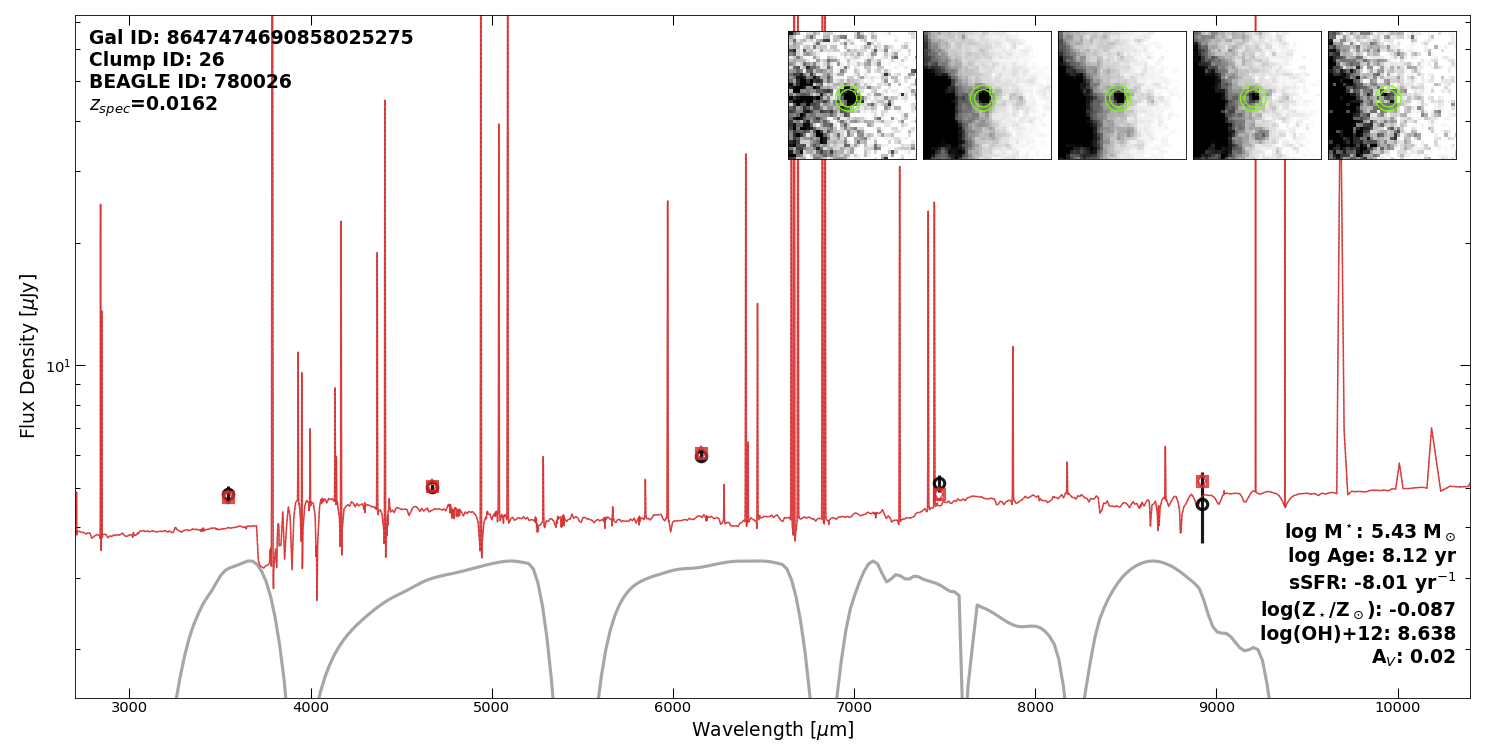}
\caption{An example illustrating the SED fit for one of the clumps identified in SDSS objID\#8647474690858025275 (same object as in Figure~\ref{fig:clump_finder} \& \ref{fig:clump_membership}). The measured clump fluxes are shown as \textit{black circles}, while the predicted fluxes for the best-fit stellar population model are shown in \textit{red squares}. The high-resolution spectrum of the best-fit stellar population model is also shown in \textit{red}. The individual $ugriz$-band stamps (\textit{from left to right}) for the clump are shown as insets in the \textit{top right} corner. The \textit{gray curves} shown are the response curves for the SDSS-$ugriz$ filters. See Section~\ref{sec:beagle} for further details on the SED fitting procedure.}
\label{fig:sed_example}
\end{figure*}

\subsection{Determining the clump galactocentric distances}
\label{sec:gal_props}

One of the critical scientific goals of this work is to investigate gradients in clump physical properties as a function of their distance from the galactic center. Hence, it is important to have a proper measurement of the galactocentric distance for the clumps. Determining the galaxy center is not a trivial process when the galaxy morphologies are chaotic and there are bright clumps residing within the galaxy.

Since we now have the knowledge of all the clumps within any host galaxy, we remeasure the CoM for each galaxy by masking all the identified clumps. However, to ensure that the bulge itself is not masked due to being misidentified as a clump, we implement an iterative process where the CoM is recomputed by unmasking any clumps within the seeing FWHM until the CoM position converges to within 1 pixel (0.396\arcsec). We also remeasure the morphological parameters -- effective radii, semi-major and semi-minor axes, and position angle -- for all host galaxies in order to account for the detected clumps. The ellipse parameters (semi-major and semi-minor axes, and position angle) are computed using the second moments of the images after masking all detected clumps and neighboring objects. For computing the effective radius, we first create a radial light profile of the galaxy where the pixels associated to clumps are masked and replaced with the average flux computed in an annulus at the masked pixels' radius. The effective radius is then computed as the circular radius that encompasses 50\%\ of the total galaxy flux.

\subsection{Stellar Population Analysis}
\label{sec:beagle}

We model the spectral energy distribution (SED) for the 543 clumps or masses in our sample using the \beagle\ tool \citep[BayEsian Analysis for GaLaxy sEds, version 0.24.0; ][]{beagle}, which allows for a self-consistent fitting of the stellar and nebular components. The Bayesian framework of \beagle\ also allows for proper quantifications of the errors on the best-fit parameters of the stellar population. The $ugriz$ photometry of the clumps after subtracting the diffuse galaxy light component is used for performing the SED fitting. Figure~\ref{fig:sed_example} shows an example of the best-fit SED for one of the clumps in our sample.

The inclusion of nebular emission lines in the SED fitting process is critical for young, star-forming stellar populations, which is expected for the clumps in this work. The implemented \beagle\ version adopts photoionization models of SFGs from \citet{gutkin16}, which combine the intrinsic emission spectrum from stars using the latest version of the \citet{bc03} stellar population synthesis models with the nebular emission from photoionized interstellar gas modeled using the radiative transfer code, \cloudy\ \citep{ferland13}. We consider a \citet{chabrier03} initial mass function with an upper mass cutoff of 100 M$_\odot$ and a fixed hydrogen density of 10$^2$ cm$^{-3}$ in the ionized gas.

For the analysis of the clump stellar population properties, the \beagle\ SED fitting is parameterized assuming a constant star-formation history (SFH) allowing the stellar mass, star-formation rate (SFR), stellar population age, metallicity, and effective dust attenuation as free parameters. We adopt the \citet{cf00} two-component dust prescription for modeling the dust. The redshift is held fixed at the redshift of the host galaxy as reported in the SDSS catalog. We specifically set the timescale for the episode of star-formation to 1 Myr to allow for the possibility of extremely young, star-bursting stellar populations, which may be expected for the clumps we are considering. Moreover, we constrain the range of stellar metallicities over a range of $\log{Z/Z_\odot}=[-1.52,0.01]$ based on the observed metallicities of clumps from their spectra (see Section~\ref{sec:clump_metallicity} for details).

\subsection{Clump Catalog}
We have compiled a catalog with all measured photometry and stellar population properties of the clumps. We also remeasure the photometry for the host galaxies to be consistent with the analysis in this paper and to obtain accurate stellar masses. For estimating the stellar population properties for the host galaxies, we follow the same algorithm as described in Section~\ref{sec:beagle} with two changes: we assume an exponentially declining SFH and allow for a wider range of stellar metallicities ($\log{Z/Z_\odot}=[-2.2,0.25]$). The full catalog is available in electronic format online\footnote{The clump catalog is available from \url{https://z.umn.edu/GZH_S82_Clumps}} and Table~\ref{tab:coldefs} defines the various columns included therein.

\section{Results}
\label{sec:results}

In this section, we present and discuss the primary results from the analysis of the clumps identified in the Stripe 82 galaxies.

\begin{figure}[!h]
\centering
\includegraphics[width=0.48\textwidth]{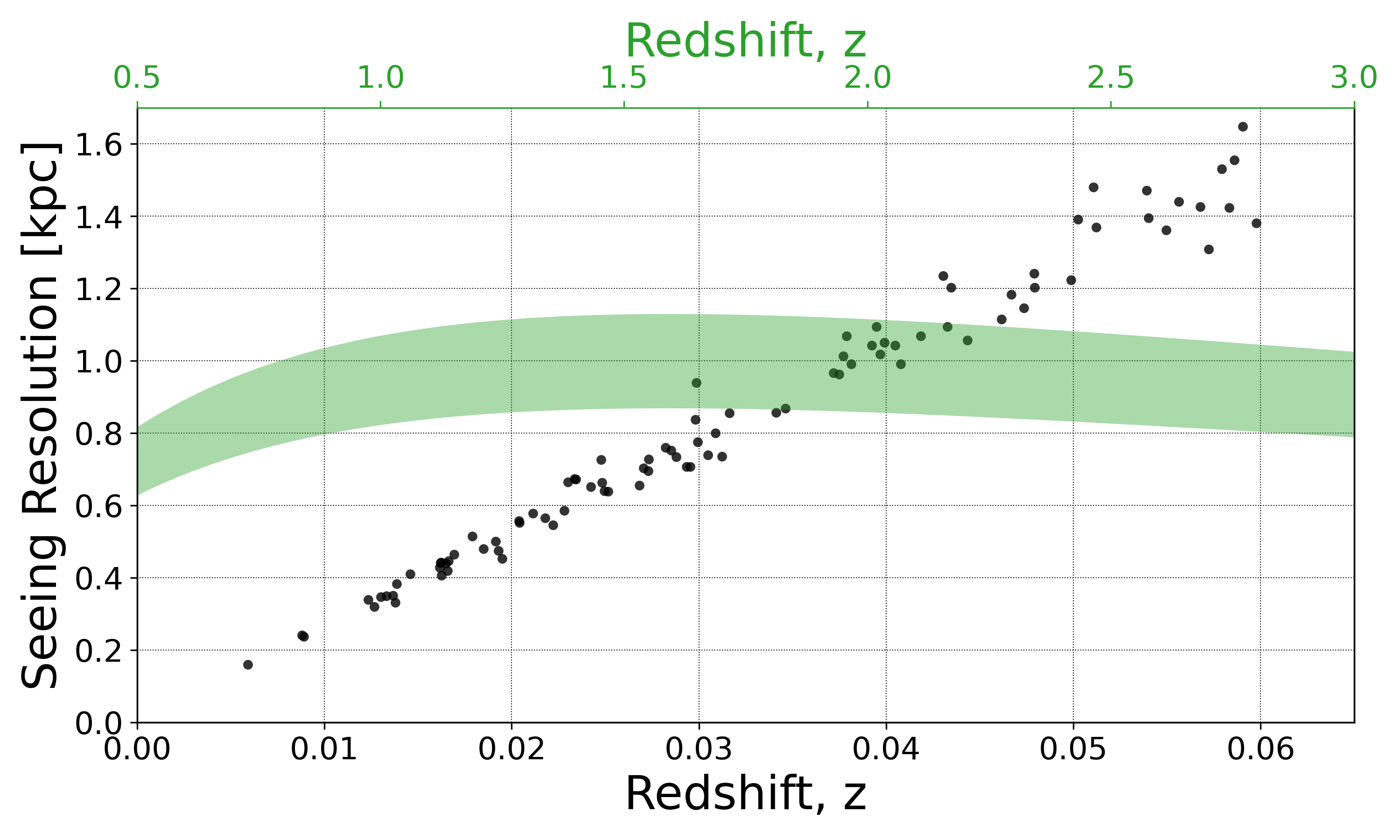}
\caption{Physical resolution probed by the SDSS-based imaging of the Stripe 82 clumpy galaxies studied in this work shown as \textit{black points}. The resolution is limited by the seeing during the observations; here, we use the $r$-band PSF FHWM of the individual galaxies to illustrate the resolving power. For comparison, the HST resolution at $0.5<z<3$ is also shown in \textit{green} on a twinned axis. Here, we estimate the HST PSF FWHM as 0.1\arcsec-0.13\arcsec.}
\label{fig:clump_seeing}
\end{figure}

\begin{figure*}[!ht]
\centering
\includegraphics[width=0.85\textwidth]{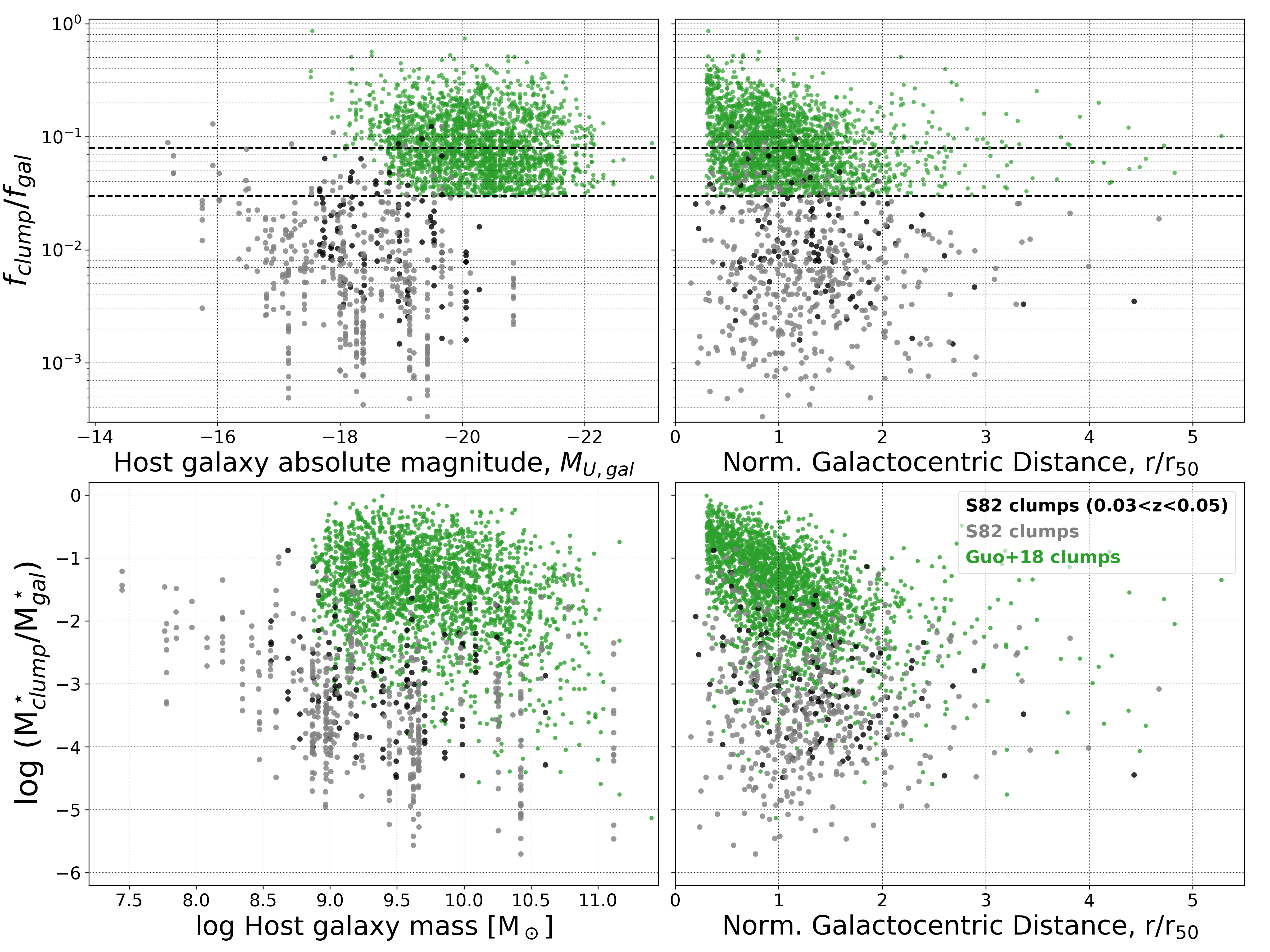}
\caption{Fractions of galaxy light and stellar mass contained within the Stripe 82 clumps. The \textit{top row} shows the fraction of the total galaxy light contained within individual clumps as a function of the total galaxy absolute $u$-band magnitude (\textit{left panel}) and the clumps' galactocentric distances normalized by the galaxy's effective radius (\textit{right panel}). For the $z\sim0$ Stripe 82 clumps (\textit{black} points for the full sample and \textit{gray} points for $0.03<z<0.05$), the fraction of flux in the $u$-band is shown; whereas for \citet{guo18} $0.5<z<3$ clump sample (\textit{green} points), the fraction of UV flux is shown. The two \textit{black dashed lines} denote 3\%\ and 8\%\ flux fractions. The \textit{bottom row} shows the fraction of the total galaxy stellar mass contained within the individual clumps as a function of the galaxy stellar mass (\textit{left panel}) and the clumps' galactocentric distances (\textit{right panel}).}
\label{fig:clump_frac_ML}
\end{figure*}

\subsection{Clump properties}

One of the primary motivations of this study is to investigate the evolution in clump properties over redshift, specifically those of the $z\sim0$ clumps found in this study compared to the clumps samples at $0.5<z<3$ from \citet{guo15,guo18}. Since both the surveys cover a different range of redshift and utilize observations from different instruments, the first challenge is to account for the difference in imaging resolution and consequently, the physical scales probed by the two surveys. We specifically target the $z<0.06$ redshift space for this SDSS-based study because the physical scales resolved by the SDSS imaging are comparable to that of HST from the \citet{guo18} clump analysis at $0.5<z<3$. Figure~\ref{fig:clump_seeing} illustrates the physical resolution of the Stripe 82 galaxies covered with $\sim1.2\arcsec$ seeing resolution alongside the $\sim0.1-0.13\arcsec$ resolution available from HST over $0.5<z<3$ and as evident, the two are comparable over $0.03<z<0.05$. In the following analysis (and figures), we highlight the $0.03<z<0.05$ clump sample for clarity; however, the qualitative differences in properties of clump within this redshift range and the full Stripe 82 clump sample are minimal and our conclusions are not significantly impacted if we were to explicitly restrict the analysis to $0.03<z<0.05$.

Before drawing comparisons, it is essential to confirm that our clump sample is qualitatively similar to the clumps found in the \citet{guo15,guo18} studies. The definition of a clump has been inconsistent throughout the literature. While earlier clump studies relied on defining clumps by visual classification \citep[e.g.,][]{elmegreen05a,elmegreen07,puech10}, follow-up studies applied detection algorithms that were made robust against changes in resolution and depth \citep[e.g.,][]{livermore12,guo12,tadaki14,zanella19}; however, a direct comparison is still nontrivial.

The \citet{guo15,guo18} studies adopted a more physical clump definition based on the fraction of total galaxy light contained in the clumps. In \citet{guo15}, only clumps with at least 8\%\ of the total rest-UV ($\sim$2300\AA) luminosity of their galaxies were selected and later relaxed to 3\%\ in their follow-up analysis \citep{guo18}. However, their definitions are specifically optimized for the detection and consistent study of high-redshift clumps. Other recent works use a similar criterion \citep[e.g.,][]{shibuya16} or have adapted the flux-fraction definition for their work -- e.g., \citet{fisher17} who instead used a fraction of \ha-flux. \citet{huertas20}, on the other hand, used a pure mass cut ($M_{clump}>10^7$ M$_\odot$) to define their clump sample in their machine learning-based study.

In our case, we perform a visual inspection of the clumps and confirm that even for cases where the fraction of total $u$-band ($\sim$3600\AA) flux is $<3\%$ (bluest band available for our sample), the recovered clumps are real and verifiable. Hence, we choose not to use any flux cut for our sample and instead, select all clumps that are identified by the contrast-based detection algorithm and visual vetting process, regardless of what fraction of the galaxy's flux is contained within.

As shown in Figure~\ref{fig:clump_frac_ML}, the clump-to-galaxy $u$-band flux ratio distribution extends down to a few$\times$0.01\%\ for our sample. We are confident in the purity of our sample for several reasons. Firstly, the \citet{guo18} cut is performed using rest-UV fluxes, whereas we are comparing $u$-band flux ratios as this is the bluest band available for our sample and the difference between a star-forming and an older stellar population is less significant at rest-$u$ ($\sim$3600\AA) compared to rest-UV ($\sim$2300\AA). Second, the clump identification for our sample is based on a $ugri$ multi-band detection and hence it is not solely reliant on the $u$-only flux ratio. Thirdly, we visually verify each clump and confirm that it is real. We have additional checks to ensure that any foreground stars and background galaxies that have been detected independently in SDSS (either photometrically or have a spectrum) are removed. Lastly, existing clumps studies in the local universe also confirm clump-to-galaxy flux ratios that extend well below the \citet{guo18} limit -- e.g., the clump-to-galaxy \ha\ flux (another strong star-formation indicator) ratio for DYNAMO galaxies ($0.06<z<0.15$) that is as low as $\sim0.7\%$ \citep{fisher17}.

Similar to the flux ratio, we also compare the ratio of the clump-to-galaxy stellar mass. As shown in Figure~\ref{fig:clump_frac_ML} (\textit{bottom row}), the Stripe 82 clumps identified in this work are on average lower mass compared to the \citet{guo18} $0.5<z<3$ clumps; however, they also reside in host galaxies that are lower mass and fainter by comparison.

\begin{figure}[!h]
\centering
\includegraphics[width=0.48\textwidth]{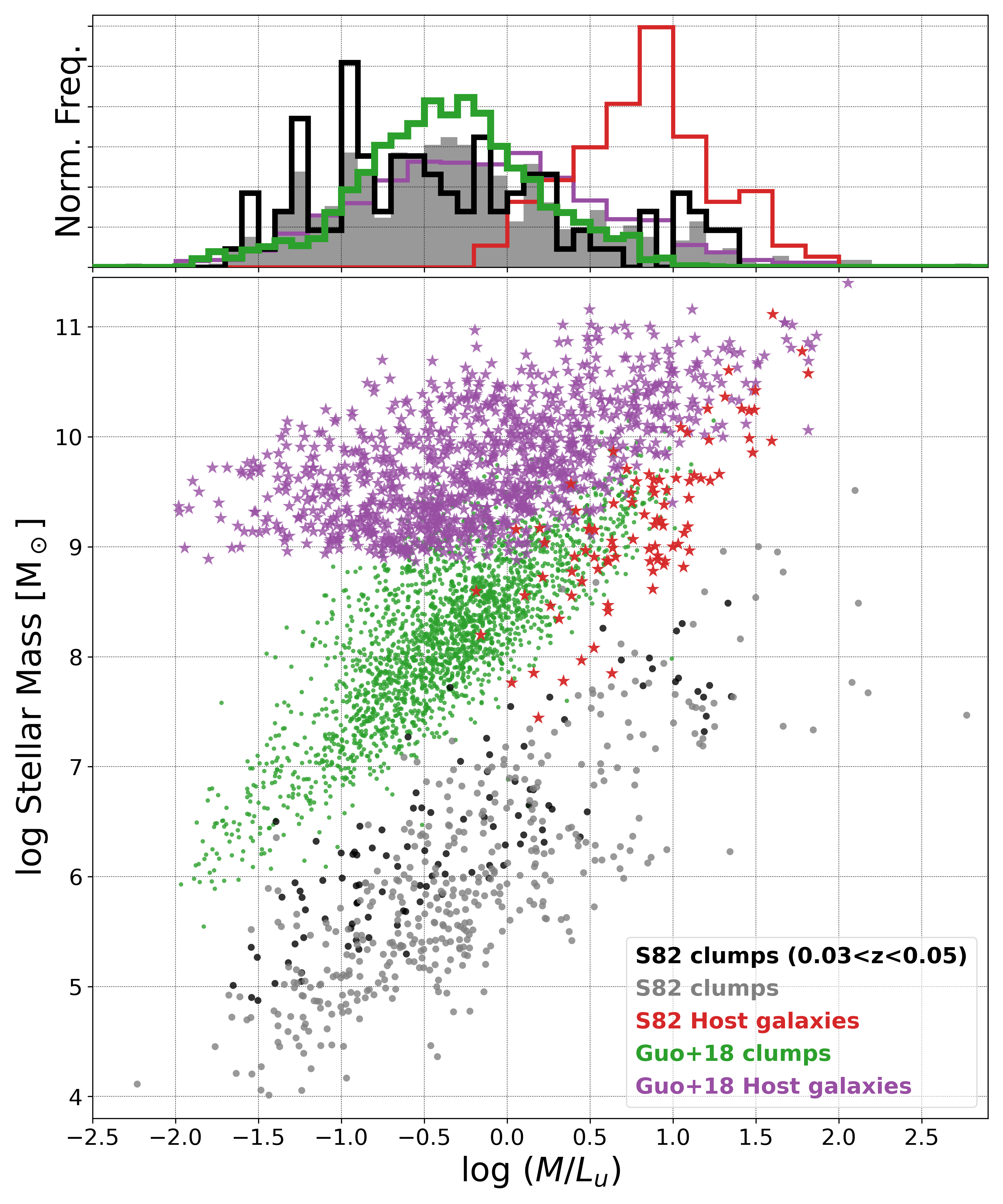}
\caption{The mass-to-light ratio for the Stripe 82 clumps shown alongside their host galaxies. The \textit{bottom panel} shows as a function of the mass-to-light ratio (computed using the $u$-band fluxes for all samples), the stellar mass for the Stripe 82 clumps (\textit{black} and \textit{gray} points) as well as their host galaxies (\textit{red} stars). For comparison, the $0.5<z<3$ \citet{guo18} clumps are also shown (in \textit{green} points) along with their host galaxies (\textit{purple} stars). The \textit{top panel} shows a normalized distribution of the mass-to-light ratio for each of the samples.}
\label{fig:clump_mass_to_light}
\end{figure}

A more direct way to compare the stellar populations contained within the clumps is to compare the mass-to-light ratio. Figure~\ref{fig:clump_mass_to_light} shows the mass-to-light ratios computed using $u$-band fluxes for our sample along with the \citet{guo18} clump sample. The \citet{guo18} mass-to-light ratios shown are computed using the filter that most closely matches the SDSS $u$-band. As evident from Figure~\ref{fig:clump_mass_to_light}, while the masses of the individual clumps identified in this study are on average lower than the $0.5<z<3$ clump sample from \citep{guo18}, the mass-to-light ratio distributions are consistent. On the other hand, there is a significant evolution in the mass-to-light ratios of the galaxies that host these clumps between $z\sim2$ and $z\sim0$. The consistency in the clump mass-to-light distributions as well as colors (see Section~\ref{sec:color_mass}) across redshift confirms that the stellar populations that are contained within the $z\sim0$ and $z\sim2$ clumps are similar.

Also shown in Figure~\ref{fig:clump_frac_ML} (\textit{right column}) are distributions of the clump-to-galaxy flux and stellar mass ratios as a function of the clumps' distance from the galactic center. This provides a check against potential systematics in our automated clump finder technique. The clump fluxes and masses recovered as a function of galactocentric distances span across the full range of clump properties from \citet{guo18} and do not show any systematic biases or selection effects, with the exception of a lack of bright clumps in the S82 sample, which is a consequence of the redshift evolution in the mass-to-light ratios of galaxies.

\begin{figure}[!h]
\centering
\includegraphics[width=0.48\textwidth]{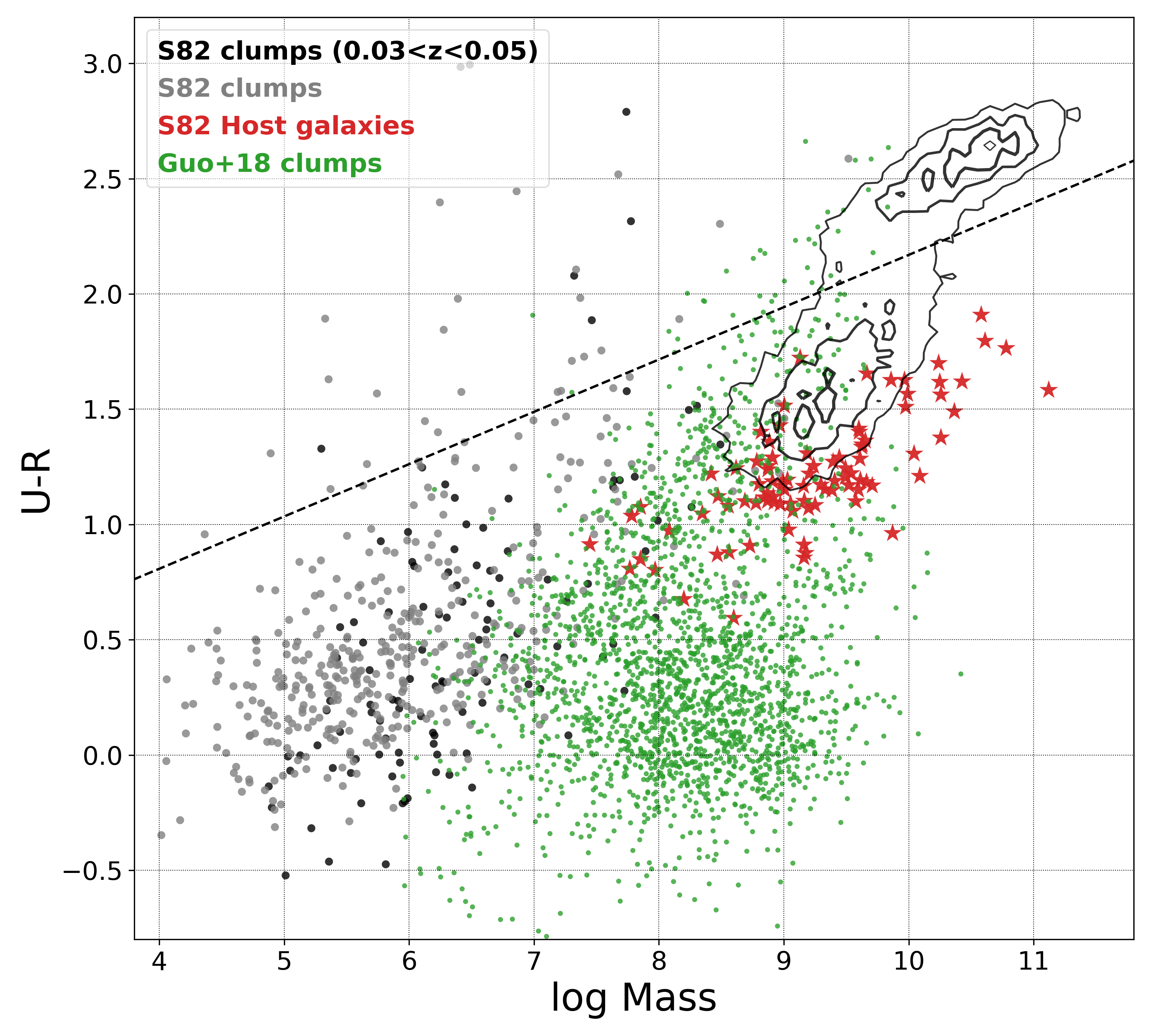}
\caption{Color-mass diagram showing the rest-frame $U-R$ color of clumps and their host galaxies as a function of their stellar masses. Stripe 82 clumps from this work are shown in \textit{black} (for $0.03<z<0.05$) and \textit{gray points} (for the full sample), whereas their host galaxies are shown as \textit{red stars}. The \citet{guo18} $z\sim1-3$ clumps are shown as \textit{green points} for comparison. The local SDSS galaxies are shown as the \textit{black contours}. The observed separation between the quiescent red sequence and the star-forming blue cloud is shown by an empirical relation plotted as the \textit{black dashed line}.}
\label{fig:clump_color}
\end{figure}

\subsection{Clumps: Color-Mass Diagram}
\label{sec:color_mass}

A color-mass diagram is an efficient way to separate star-forming and quiescent objects.  Figure~\ref{fig:clump_color} shows the Stripe 82 clumps and their host galaxies on a rest-frame $U-R$ vs. stellar mass, $M^\star$, diagram. The full SDSS $z<0.06$ galaxy sample (shown as contours) makes obvious the distinct separation in the colors of the red sequence and the blue cloud. An empirical line tracing this separation is shown for reference.

Overall clumps are on average bluer than their host galaxies, which is expected considering that these clumps are experiencing ongoing star-formation and have younger stellar populations as also confirmed in Figure~\ref{fig:clump_mass_to_light}. The colors of the clumps are comparable with the expected colors for a star-forming population if one were to extrapolate the observed empirical relation (that separates the star-forming and quiescent populations) down to lower masses. While some of the Stripe 82 clumps have colors consistent with the quiescent galaxy population, we do not exclude them from the sample so as to not exclude the possibility of \textit{ex-situ} clumps. For comparison, the \citet{guo18} sample of high-redshift clumps is overplotted in Figure~\ref{fig:clump_color} and qualitatively, they exhibit the same behavior.

\begin{figure}[!h]
\centering
\includegraphics[width=0.48\textwidth]{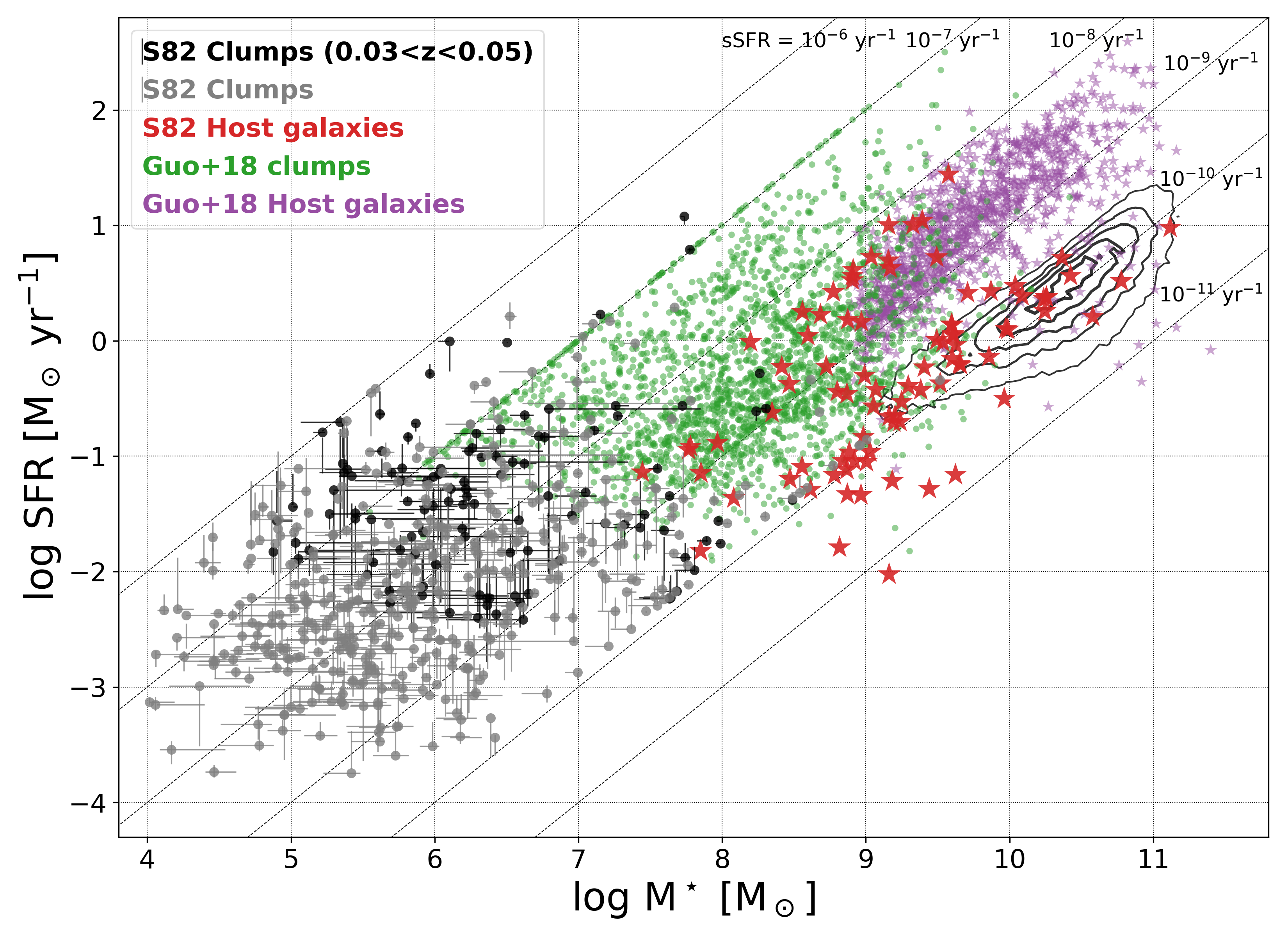}
\caption{The SFRs of clumps as well as their host galaxies plotted as a function of their stellar masses. The clumps found in the Stripe 82 galaxies in this work are plotted as \textit{black} (for $0.03<z<0.05$) and \textit{gray points} (for the full sample) and their host galaxies as \textit{red stars}. The local star-formation galaxy population is shown in \textit{black contours} to highlight the $z\sim0$ star-forming main sequence. Clumps found in $z\sim1-3$ CANDELS galaxies from \citet{guo18} are shown as \textit{green} points and their host galaxies as \textit{purple} stars. For reference, lines of constant sSFR values are also shown.}
\label{fig:clump_sfms}
\end{figure}

\subsection{Clumps: SFR vs. $M^\star$}

The star-forming main sequence (SFMS) is arguably the most prominent feature amongst star-forming galaxies (SFGs) extending from $z\sim0$ out to $z\sim8$ with a surprisingly tight correlation between the star-formation rate and stellar mass of SFGs \citep{speagle14}. The SFMS is often plotted using integrated quantities, the star-formation rate and stellar mass of a galaxy, averaged over several individual star-forming regions spanning across a galactic scale. On the other hand, the clumps studied in this work probe star-formation on a much smaller scale ($\sim$1 kpc). Hence, the position of these clumps on the SFMS can be particularly insightful.

Figure~\ref{fig:clump_sfms} shows the star-formation as a function of stellar mass for the Stripe 82 clumps as well as their host galaxies. As evident from the figure, a sizeable fraction of the Stripe 82 host galaxies reside along the $z\sim0$ SFMS as outlined by the contours that show the local SFG population. In contrast, the clumps residing within them extend across a vast range of sSFR values from 10$^{-10}$ to 10$^{-6}$ yr$^{-1}$. While the clumps themselves are significantly lower mass than their host galaxies, their sSFR is preferentially higher than the SFMS indicating that star-formation in these clumps is much more vigorous than the galactic average. This result is consistent with other clump studies \citep[e.g.,][]{guo12,wuyts12,guo18}.
Across various recent studies, it is becoming increasingly clear that star-formation in galaxies is not a secular process, but rather proceeds in stochastic bursts \citep[e.g.,][]{dominguez15,mehta17,emami19}. The presence of intensely star-forming clumps in galaxies could be one manifestation of stochastic bursts of star-formation.

For comparison, the clumps found in $z\sim1-3$ CANDELS galaxies \citep{guo18} are also plotted along with their host galaxies. The host galaxies of both the high-redshift and the local clump samples consist predominantly of galaxies that lie on the SFMS at their corresponding redshifts. Both clump samples also span a similar range of sSFR despite the high-redshift clumps being more massive on average compared to the local clumps. The high-redshift clump sample is artificially truncated at sSFR=10$^{-7}$ yr$^{-1}$ due to the minimum allowed star-formation timescale when fitting stellar populations in \citep{guo18}. Qualitatively, in terms of their star-formation properties, the local clumps studied in this work are similar to the high-redshift clumps from \citet{guo18}.

\begin{figure}[!h]
\centering
\includegraphics[width=0.48\textwidth]{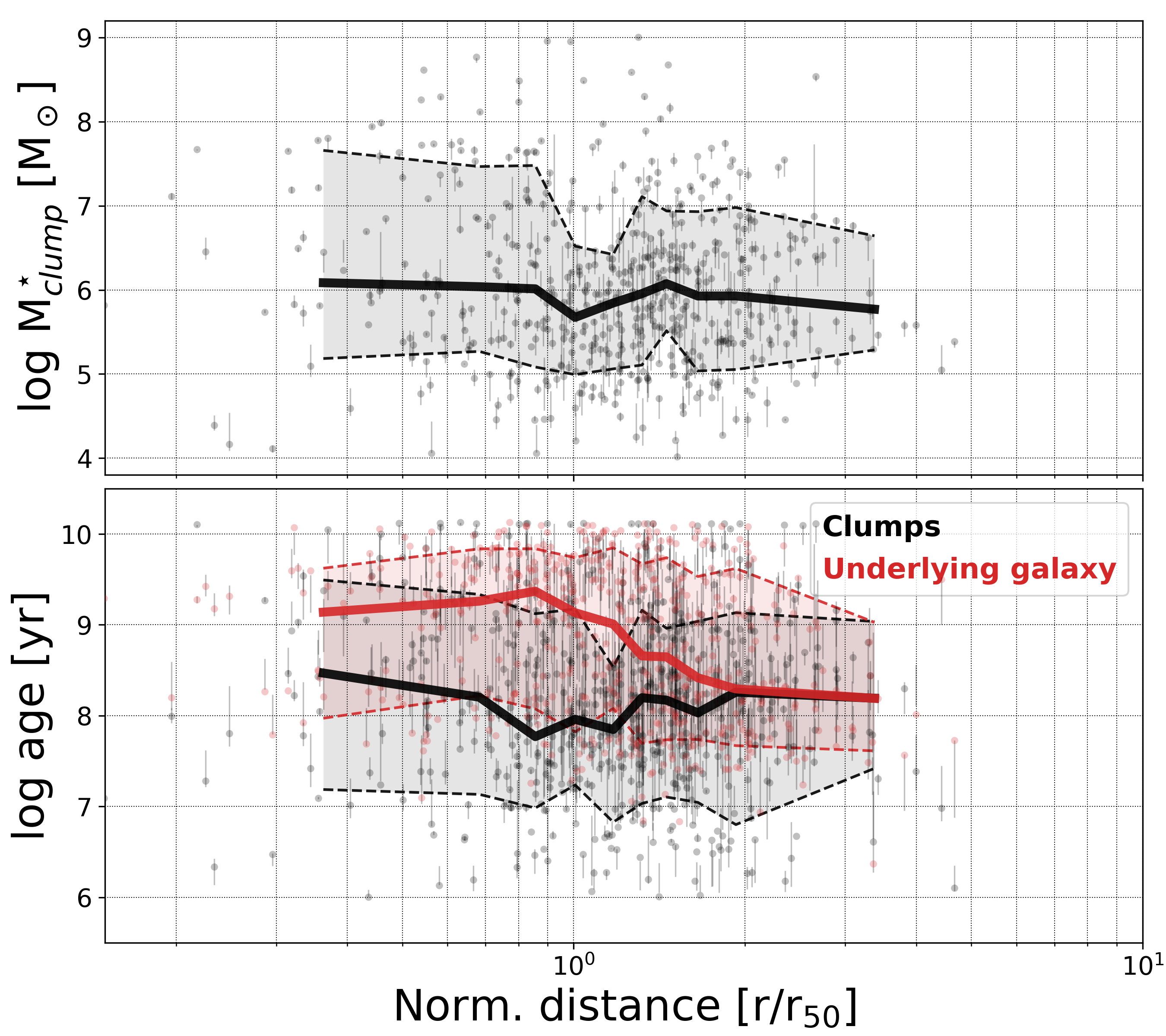}
\caption{Clump masses and ages as a function of their galactocentric distance shown in \textit{black points}. A running median is shown with the \textit{black line} with the shaded band showing the $16^{\text{\tiny th}}$ and $84^{\text{\tiny th}}$ percentile range. Similarly, the ages and mass derived from SED fitting the flux contributed by the underlying galaxy light at the clump location are shown in \textit{red points} and its running median with the \textit{red line}. See text for details.}
\label{fig:clump_gradient}
\end{figure}

\begin{figure}[!h]
\centering
\includegraphics[width=0.48\textwidth]{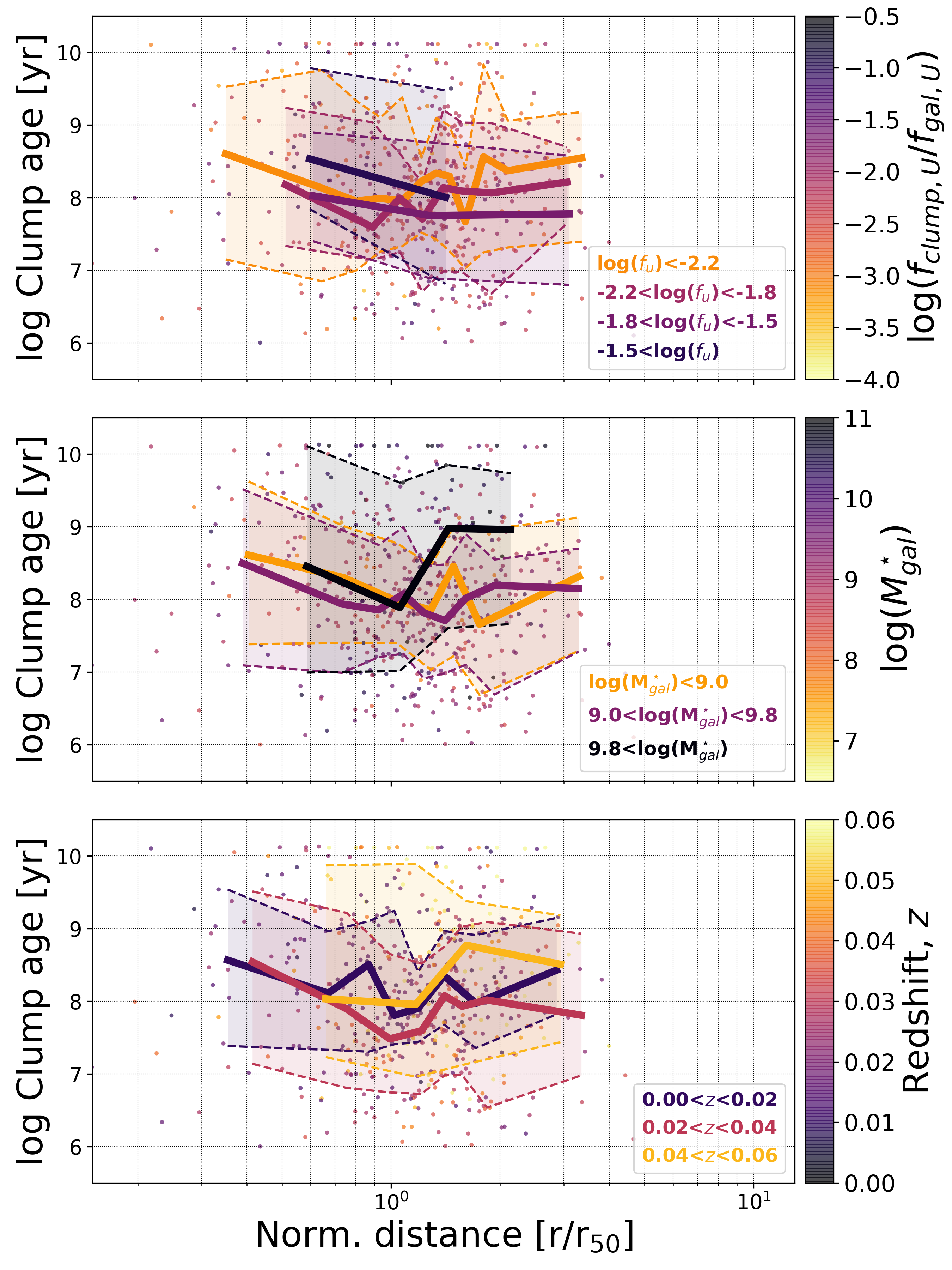}
\caption{Same as the bottom panel of Figure~\ref{fig:clump_gradient}, but now points are binned according to the fraction of a galaxy's $u$-band light contained within the clump (\textit{top} panel), host galaxy mass (\textit{middle} panel) and redshift (\textit{bottom} panel).}
\label{fig:clump_gradient2}
\end{figure}

\subsection{Clump mass and age gradients}
\label{sec:clump_gradient}

Gradients in clump ages as a function of their galactocentric distance is one of the critical tests for different theoretical models on clump evolution. Models based on inward migration of clumps \citep[e.g.,][]{bournaud07,bournaud14,elmegreen08,ceverino10,mandelker14} predict a negative age gradient -- older ages for clumps that are closer to the galactic center, and younger ages for those in the outskirts. The strength of the age gradient in these models is predicted to be on the order of a few hundred Myr, as the clumps migrate from the outskirts to the galactic centers. Observationally, studies investigating clumps at $z\sim1-3$ have confirmed a negative age gradient consistent with inward migration \citep[e.g.,][]{schreiber11,guo12,guo18,soto17}. On the other hand, some simulations \citep[e.g.,][]{buck17,oklopcic17} have argued that the age gradient is merely a result of clumps being contaminated by stars from the underlying galactic disk at the clump location.

Figure~\ref{fig:clump_gradient} shows stellar masses and ages for Stripe 82 clumps as a function of their galactocentric distance. As evident from the figure, we find no significant dependence of clump ages or masses as a function of their distance. For the clumps, the fitted stellar populations assumed a constant SFH. We refit our sample assuming an exponential SFH and confirm that the result is not impacted by the choice of the SFH when fitting SEDs. Furthermore, in Appendix~\ref{appndx:sed_degeneracy_test}, we confirm that our result is not affected by degeneracies in the estimated stellar age, dust extinction, and metallicity that may be introduced during SED fitting.

Since there is only $ugriz$ photometric band information available for our clumps, one obvious systematic check to perform is whether the five photometric bands are sufficient to recover an age gradient, if one were present. For this test, we check the ages of the underlying galaxy light at the locations of the clumps since an age gradient in the galactic disk has been observationally confirmed \citep[e.g.,][]{gonzalez14,gonzalez15,liu16,ruiz17,wang17,tacchella18} and is to be expected for our sample of SFGs.

The flux from the galactic disk at the location of the clumps is computed as part of the analysis and readily available, since the clump photometry is treated for contamination from underlying galaxy light (see Section~\ref{sec:photometry}). Moreover, these apertures will be affected by the same sampling biases as the sample of real clumps, in terms of their host galaxy properties, thus providing a more direct comparison. We fit the galaxies' flux contributions to the clump apertures using the same SED fitting procedure with one exception -- we assume exponentially declining SFHs, which are more appropriate for a typical star-forming galaxy.

As shown in Figure~\ref{fig:clump_gradient}, the stellar population properties of the galaxy light at the clump locations reveal a clear gradient in their ages as a function of distance. Thus, the expected gradient in galactic age has been successfully recovered with the $ugriz$-band information available for our sample. Critically, this test demonstrates that our data are clearly capable of recovering an age gradient, if one were present. Moreover, the discrepancy in the fitted clump and galactic disk ages also verifies that the diffuse galaxy light subtraction performed for clump photometry has a clear impact on the recovered clump ages. In Appendix~\ref{appndx:age_gradient_test}, we extend this test using the full galaxy disk instead of just the galaxy light at the clump locations. In conclusion, we find no significant gradient detected in the stellar ages for our sample of clumps in local Stripe 82 galaxies.

The lack of a gradient in the clumps' ages challenges the scenario of clumps forming \textit{in-situ} and migrating inwards as they evolve, which has been suggested from clump studies at high redshift \citep[$z\sim1-2$; e.g.,][]{soto17,guo18}. Our findings suggest that there could be multiple factors at play, instead of a single channel for clump formation and evolution in the local universe. Firstly, the clumps in the outskirts could include some that have been formed \textit{ex-situ} with intrinsically older ages \citep[e.g.,][]{mandelker14}. The presence of these \textit{ex-situ} clumps could smear or flatten the age gradient. It is also likely that the \textit{ex-situ} contribution to clump formation is larger at lower redshifts given the higher frequency of minor mergers \citep{lotz11,bluck12,guo15}.

Secondly, while the migration scenario is disfavored, we cannot fully rule it out. The clump ages in our sample are much older than those expected for the short-lived clump scenario \citep[$\lesssim$50 Myrs; e.g.,][]{oklopcic17}; instead, the Stripe 82 clump ages are strongly in favor of the long-lived clump scenario, which suggests that clumps survive for long periods of time \citep[e.g.,][]{bournaud07,mandelker14,mandelker17}. Whether or not the observed clumps have migrated toward the center cannot be definitely established with our current data. The typical migration timescale reported in numerical simulations is a few hundred Myr \citep{elmegreen08,ceverino10}; however, this value is computed for high-redshift ($z\sim1$) galaxies. In the local universe, where the disks have lower gas fractions than at high redshifts, the migration timescales may be different and are expected to be longer considering the increase in the dynamical timescales of galaxies as well as the stabilization of their disks. In conclusion, while our results challenge the inward migration scenario, we are unable to rule it out. A combination of factors such as larger contribution of \textit{ex-situ} clumps along with longer migration timescales in our Stripe 82 galaxy sample could also make the age gradient too flat to be detectable.

Negative age gradients have been detected in studies at $z\sim2$ that apply a clump-to-galaxy flux ratio for selecting clumps \citep[e.g.,][]{guo18}. We therefore explicitly investigate the impact of the clump-to-galaxy flux ratio on the age gradient. Figure~\ref{fig:clump_gradient2} (\textit{top} panel) shows the age gradient from Figure~\ref{fig:clump_gradient} broken up into four bins of clump-to-galaxy $u$-band flux ratios, $f_U = f_{clump,U}/f_{gal,U}$. Even for the sample of clumps with $f_U\gtrsim3\%$ cut (matching that of \citealt{guo18}), we find no evidence for a clump age gradient as a function of galactocentric distance.

Furthermore, we split the sample according to the host galaxy mass (shown in the \textit{middle} panel of Figure~\ref{fig:clump_gradient2}) to investigate any potential dependence on the host galaxy properties. For both the $\log{(M_{gal}/M_\odot)}<9$ and $9<\log{(M_{gal}/M_\odot)}<9.8$ bins, we find no significant age gradient. For the high-mass bin ($\log{(M_{gal}/M_\odot)}>9.8$), there is a slight deviation toward older clumps toward larger radii ($r/r_{50}\gtrsim1.25$). This could hint at a more significant contribution from \textit{ex-situ} clumps for the higher mass host galaxies.

Lastly, we also bin the sample according to the redshift of the host galaxy (shown in the \textit{bottom} panel of Figure~\ref{fig:clump_gradient2}) and confirm that the differences in the physical scales probed over the redshift range of our sample also do not impact the overall age gradient.

\begin{figure}[!h]
\centering
\includegraphics[width=0.45\textwidth]{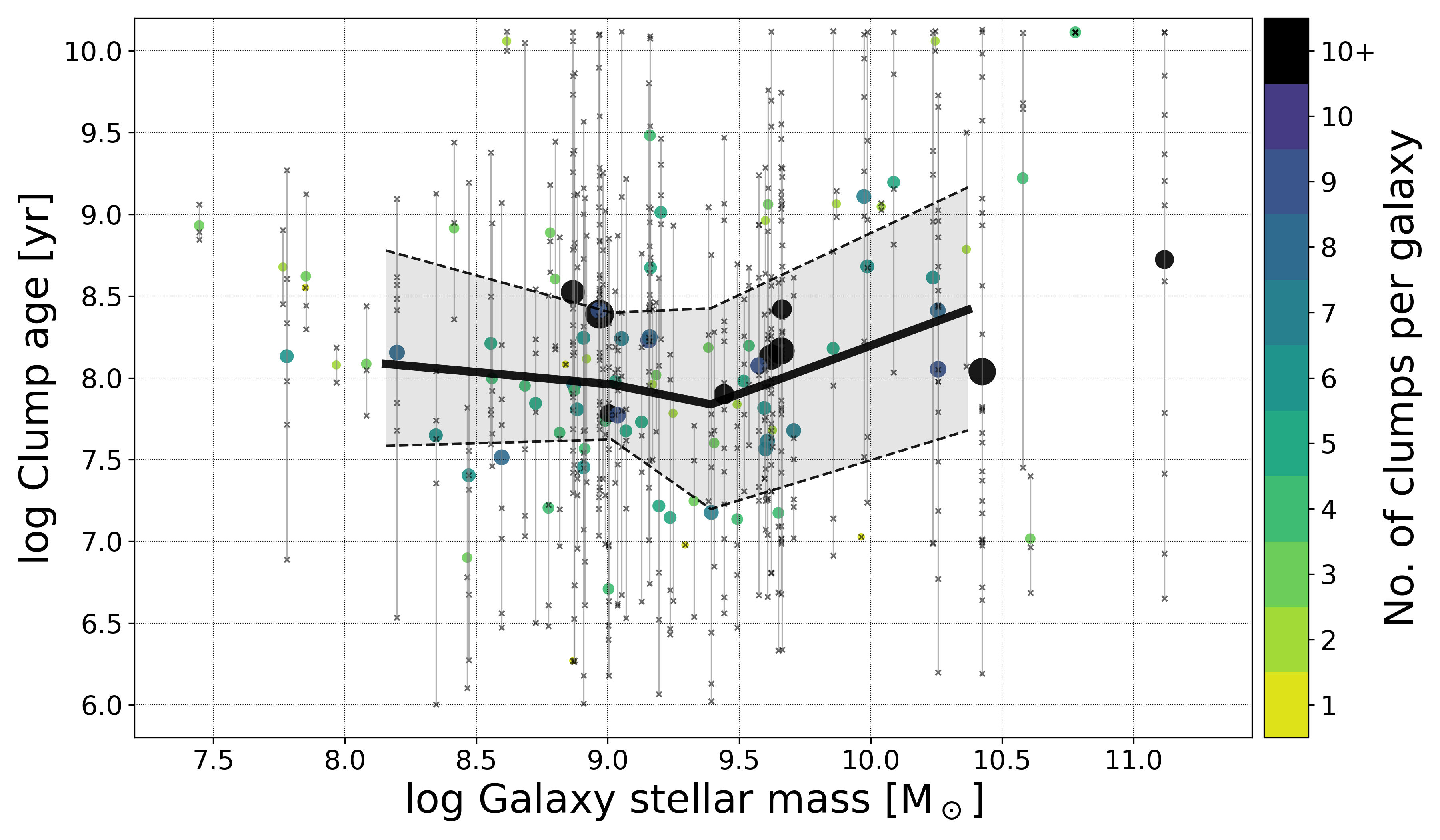}
\caption{Stellar ages of clumps shown as a function of their host galaxy mass. Each \textit{vertical line} denotes the full range of clump stellar ages (shown as \textit{crosses}) for all clumps residing in the corresponding host galaxy. The \textit{circle} symbols denote the average age of all clumps in a given galaxy, whereas the color and the size of the symbols denotes the number of clumps in a given galaxy. The \textit{black solid line} shows a running median for the average clump ages as a function of the host galaxy mass.}
\label{fig:clump_age_galmass}
\end{figure}

\subsection{Clump ages vs. host galaxy mass}

Since clumps reside within galaxies, it is reasonable to expect the clump properties to be impacted by the host galaxies' properties. Strikingly, however, we find that the stellar ages of the clumps do not depend on the mass of their host galaxy. Figure~\ref{fig:clump_age_galmass} shows the average ages of clumps residing in a given galaxy as a function of their host galaxies' stellar mass and we find no evidence of a correlation or dependence. At face value, this implies that the mass of the host galaxy does not have any impact on when the clumps form and dissipate. In other words, the clump formation mechanism appears to be independent of the host galaxy's stellar content.

\begin{figure}[!h]
\centering
\includegraphics[width=0.45\textwidth]{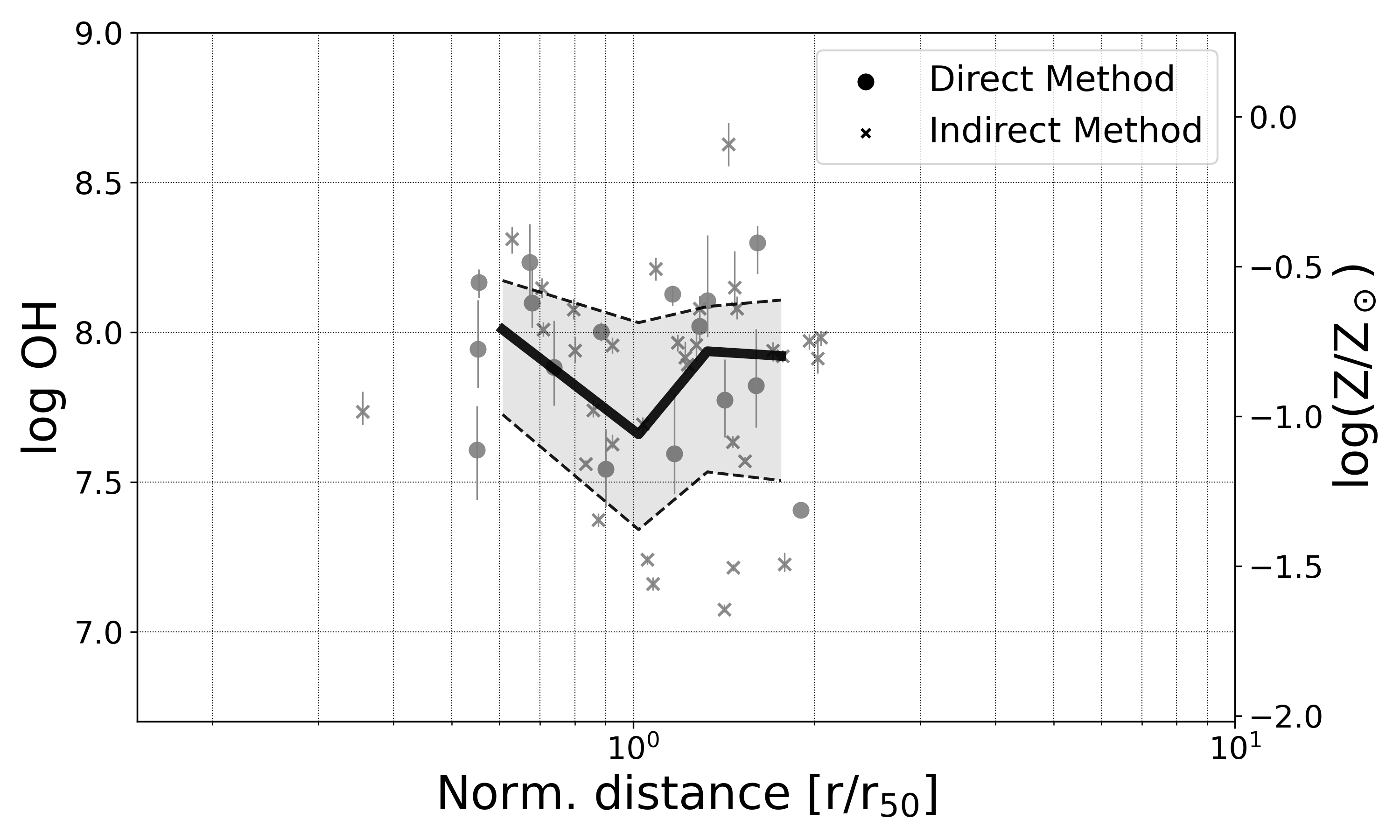}
\caption{Gas-phase metallicities for clumps measured from SDSS spectra using the direct method (where \oiiic\ is available) and indirect method (where a reliable $R_{23}$ is available). The \textit{solid black line} shows a running median. Note that the SDSS fibers include contribution from the underlying diffuse galaxy light in addition to the clump and hence, the metallicity measurement is not exclusively that of the clump.}
\label{fig:clump_metallicity}
\end{figure}

\subsection{Clump Metallicities}
\label{sec:clump_metallicity}

One of the key advantages of the local sample of clumpy galaxies is that some of the individual clumps are bright and separated enough to be detected as independent photometric sources in SDSS. A subset of these even have exclusive SDSS spectroscopy available with the spectral fiber centered at the clump location. Querying the archive for spectroscopic targets that are within 1.5\arcsec\ of identified clumps, we find 74 clumps with spectra. Of those, 49 have sufficient signal-to-noise in the emission lines required for an indirect metallicity measurement and 19 of these have a detected \oiiic\ for a direct measurement.

We manually refit the SDSS spectra to obtain a more robust measurement for the weak \oiiic\ line. For the rest of the lines, we use the SDSS measured values. We require a signal-to-noise of $>10$ for the \oiia,\oiib,\oiiia,\oiiib\ and \hb\ emission lines in order to measure the metallicity. The emission line measurements are treated for galactic extinction using the \citet{schlegel98} dust maps and assuming a \citet{cardelli89} dust law. Furthermore, we quantify the dust content in the host galaxy from the Balmer decrement (assuming Case B) and correct the emission line fluxes according to a \citet{calzetti00} dust law.

For the spectra where the \oiiic\ is detected at signal-to-noise of $>3$, we measure the metallicity for the clump fibers via the direct method \citep{izotov06} using the emission line analysis software, \pyneb\ \citep{pyneb}. Assuming an electron density of $10^{-2}$ cm$^{-3}$, we first compute the temperature for $O^{++}$ using the \oiiia, \oiiib, and \oiiic\ lines. The $O^{++}/H$ and $O^+/H$ abundances are then computed using the (\oiiia+\oiiib)/\hb\ and (\oiia+\oiib)/\hb\ line ratios, respectively. Lastly, the total oxygen abundance is computed as $O/H = O^{++}/H + O^+/H$. For spectra where the \oiiic\ is too weak to be detected, we estimate the metallicity using the indirect method. Specifically, we use the $R_{23}$ calibration from \citet{pilyugin05}. We do not expect the metallicities for the star-forming clumps to be super-solar and hence, we use the lower branch calibration.

Figure~\ref{fig:clump_metallicity} shows the direct and indirect metallicity measurements for the Stripe 82 clumps, where possible. Overall, we find no significant gradient in the metallicity as a function of distance for these local clumps. However, a word of caution when interpreting this result -- the SDSS spectral fiber has an aperture of 3\arcsec\ (2\arcsec\ in the case of BOSS spectra) and hence, the diffuse galaxy light is expected to contribute and contaminate the clump spectra. Since there is no straightforward way to separate the clump and galaxy flux contributions to the spectra (as opposed to the case with photometry), the metallicities measured from the spectra are likely to be affected by contamination from galaxy light.

\startlongtable
\begin{deluxetable*}{lll}
\tablewidth{\textwidth}
\tablecaption{Column Descriptions for the clump catalog \label{tab:coldefs}}
\tabletypesize{\scriptsize}
\tablehead{
\colhead{Column No.} &
\colhead{Column Title} &
\colhead{Description} }
\startdata
                 \multicolumn{3}{c}{\textbf{Galaxy information}}\\
                 \hline
                 1       & id                             & Galaxy Identification number in SDSS Stripe 82 coadd-epoch catalog \\
                 2       & ra                             & Right Ascension of the galaxy CoM [deg] \\
                 3       & dec                            & Declination of the galaxy CoM [deg] \\
                 4       & redshift                       & Galaxy redshift available from DR14 \\
                 5       & specobjid                      & Identification number for SDSS spectrum centered on the galaxy, if available \\
                 6       & run                            & SDSS run number \\
                 7       & rerun                          & SDSS rerun number \\
                 8       & camcol                         & SDSS camera column \\
                 9       & field                          & SDSS field number \\
                 10      & obj                            & SDSS object ID within the field \\
                 11      & psf\_fwhm\_u                   & $u$-band PSF FWHM [arcsec] \\
                 12      & psf\_fwhm\_g                   & $g$-band PSF FWHM [arcsec] \\
                 13      & psf\_fwhm\_r                   & $r$-band PSF FWHM [arcsec] \\
                 14      & psf\_fwhm\_i                   & $i$-band PSF FWHM [arcsec] \\
                 15      & psf\_fwhm\_z                   & $z$-band PSF FWHM [arcsec] \\
                 16      & gal\_reff                      & Galaxy effective radius ($R_{eff}$) [arcsec] \\
                 17      & gal\_sma                       & Galaxy semi-major axis [arcsec] \\
                 18      & gal\_smb                       & Galaxy semi-minor axis [arcsec] \\
                 19      & gal\_theta                     & Galaxy position angle [deg] \\
                 \hline
                 \multicolumn{3}{c}{\textbf{Galaxy photometry\tablenotemark{\tiny a}}}\\
                 \hline
\multirow{4}{*}{19-39}   & gal\_flux\_*                   & Flux within a Kron-like elliptical (AUTO) aperture [$\mu$Jy] \\
                 \quad   & gal\_fluxerr\_*                & Error for AUTO flux [$\mu$Jy] \\
                 \quad   & gal\_mag\_*                    & Kron-like elliptical (AUTO) aperture magnitude [AB] \\
                 \quad   & gal\_magerr\_*                 & Error on AUTO magnitude [AB] \\
                 40      & gal\_extinct\_ebv              & Galactic extinction in $E(B-V)$ at the galaxy position \\
                 41      & gal\_extinct\_av               & Galactic extinction in $A_V$ at the galaxy position \\
                 \hline
                 \multicolumn{3}{c}{\textbf{Galaxy stellar population properties\tablenotemark{\tiny b}}}\\
                 \hline
                 42-44   & gal\_logMst                    & Log stellar mass [$M_\odot$] \\
                 45-47   & gal\_logage                    & Log stellar age [yr] \\
                 48-50   & gal\_logtau                    & Log star-formation timescale (for exponential SFH) [yr] \\
                 51-53   & gal\_logSFR                    & Log star-formation rate [$M_\odot yr^{-1}$] \\
                 54-56   & gal\_Z\_star                   & Log stellar metallicity [$Z_\odot$] \\
                 57-59   & gal\_logOH                     & Gas-phase metallicity [log(OH)+12] \\
                 60-62   & gal\_AV                        & Attenuation at 5500\AA\ [mag] \\
                 \hline
                 \multicolumn{3}{c}{\textbf{Clump Information}}\\
                 \hline
                 63      & clump\_id                      & Clump identification number \\
                 64      & clump\_ra                      & Clump Right Ascension [deg] \\
                 65      & clump\_dec                     & Clump Declination [deg] \\
                 66      & clump\_specobjid               & Identification number for SDSS spectrum centered on the clump, if available \\
                 67      & clump\_prox\_flag              & Proximity flag for the clump \\
                 68      & clump\_distance                & Clump galactocentric distance [arcsec] \\
                 69      & clump\_distnorm                & Clump galactocentric distance normalized by the galaxy effective radius ($R_{eff}$) \\
                 70      & clump\_dist\_sma               & Clump galactocentric distance normalized by the galaxy semi-major axis \\
                 71      & clump\_distphys                & Clump galactocentric distance [kpc] \\
                 \hline
                 \multicolumn{3}{c}{\textbf{Clump photometry\tablenotemark{\tiny a}}}\\
                 \hline
\multirow{4}{*}{72-91}   & clump\_flux\_*                 & Flux within a Kron-like elliptical (AUTO) aperture [$\mu$Jy] \\
                 \quad   & clump\_fluxerr\_*              & Error for AUTO flux [$\mu$Jy] \\
                 \quad   & clump\_mag\_*                  & Kron-like elliptical (AUTO) aperture magnitude [AB] \\
                 \quad   & clump\_magerr\_*               & Error on AUTO magnitude [AB] \\
                 \hline
                 \multicolumn{3}{c}{\textbf{Clump stellar population properties\tablenotemark{\tiny b}}}\\
                 \hline
                 92-94   & clump\_logMst                    & Log stellar mass [$M_\odot$] \\
                 95-97   & clump\_logage                    & Log stellar age [yr] \\
                 98-100  & clump\_logSFR                    & Log star-formation rate [$M_\odot yr^{-1}$] \\
                 101-103 & clump\_Z\_star                   & Log stellar metallicity [$Z_\odot$] \\
                 104-106 & clump\_logOH                     & Gas-phase metallicity [log(OH)+12] \\
                 107-109 & clump\_AV                        & Attenuation at 5500\AA\ [mag] \\
\enddata
\tablenotetext{a}{Each filter ($ugriz$) has its own entry. The columns ending in \texttt{*} are suffixed by the filter name.}
\tablenotetext{b}{Each stellar population property is accompanied by two columns (\texttt{*\_l68} and \texttt{*\_u68}) that present its corresponding 1$\sigma$ error.}
\end{deluxetable*}

\section{Conclusions}
\label{sec:conclusions}

In this work, we present an analysis of 543 clumps identified in a sample of 92 $z<0.06$ SDSS Stripe 82 galaxies -- the first clump study in the local universe with a statistically significant sample size. We specifically restrict the redshift range of our sample where the physical resolution available from the SDSS imaging matches that of the high-redshift HST-based clump studies, allowing us to draw valid comparisons across cosmic time.

We utilize results from the citizen science-powered \textit{Galaxy Zoo: Hubble} project to select a sample of clumpy galaxies and identify star-forming clumps using a contrast-based imaging analysis algorithm. We measure the photometry of the clumps and fit stellar population synthesis models to estimate the clump physical properties. Our primary results are as follows:

\begin{itemize}
    \item We find no evidence for a gradient in clump ages or masses as a function of the clumps' galactocentric distances. This is contrary to what has been observed for high-redshift clumps \citep{guo18}. Even after applying cuts on the clump fractional flux, host stellar masses or redshift ranges to more precisely match the \citet{guo18} sample, we do not find any significant gradient in the clump ages. Our results challenge the inward migration scenario for clump evolution for the local universe, potentially suggesting a larger contribution of \textit{ex-situ} clumps and/or longer clump migration timescales.
    \item We verify and confirm that the quality of our data is sufficient to recover an age gradient for the photometry of the galactic light underlying the clumps. Hence, if a gradient in the clump ages was present, it should have been recovered.
    \item The clump sSFR values range from $10^{-10}$ yr$^{-1}$ to $10^{-6}$ yr$^{-1}$, which is in agreement with the high-redshift clumps. Similarly, the color-mass diagram confirms that the sample of local clumps is largely consistent with the star-forming galaxy population.
    \item We find that the stellar ages of clumps in a given galaxy are not significantly dependent on their host galaxy's mass. At face value, this implies that the clump formation mechanism appears to be independent of the host galaxy's stellar mass.
    \item We measure the gas-phase metallicities for 74 clumps where an exclusive SDSS spectrum was available. While the spectrum is contaminated by the underlying galaxy, we find that the gas-phase metallicities do not show any dependence on the clumps' galactocentric distances.
\end{itemize}

In the near future, we plan to greatly expand on the sample of clumps in the local universe with the help of the \textit{Galaxy Zoo: Clump Scout} project (Adams et al., in preparation), which utilizes crowd-sourced aggregation to identify clumps in a sample of over 50,000 SDSS galaxies, thus, providing a valuable and independent check on our results.

\acknowledgments
V.M., C.S., L.F., H.D., and D.A. acknowledge support from the US National Science Foundation grant AST1716602.

\software{NumPy \citep{numpy}, SciPy \citep{scipy}, AstroPy \citep{astropy13,astropy18}, Matplotlib \citep{matplotlib}, \photutils\ \citep{photutils}, \sex\ \citep{sextractor}, \beagle\ \citep{beagle}, \pyneb\ \citep{pyneb}, SDSS SciServer, \python}

\appendix

\section{Addressing parameter degeneracies in SED-fitting}
\label{appndx:sed_degeneracy_test}

Stellar age, dust extinction, and metallicity are often degenerate when fitting galaxy spectra. Since our main results are dependent on the stellar age estimated from SDSS $ugriz$-band photometry, it is important to ensure that our results are not affected by these degeneracies. Here, we check for the possibility of a stellar age gradient being present but manifesting as a gradient in the best-fit dust extinction or stellar metallicity instead, due to the degeneracy.

\begin{figure}[!h]
\centering
\includegraphics[width=0.45\textwidth]{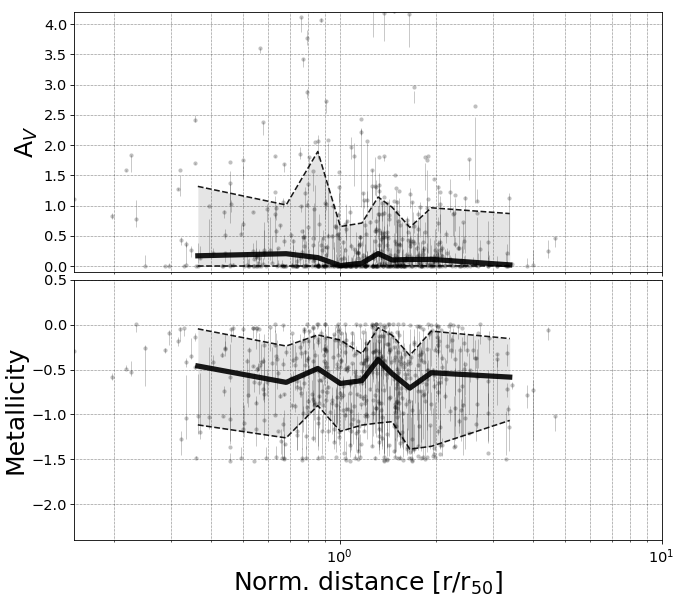}
\includegraphics[width=0.45\textwidth]{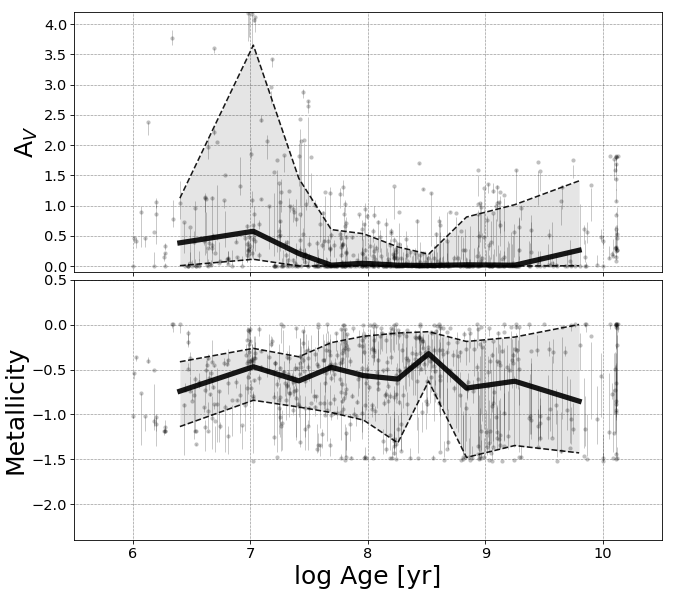}
\caption{Dust (shown as $A_V$) and metallicity ($Z^\star$) for clumps' stellar content shown as a function of the normalized galactocentric radius (\textit{left panel}) and the clump stellar age (\textit{right panel}). The solid black line shows the median value, while the dashed black lines (along with the shaded region) show the 16$^\mathrm{th}$ and 84$^\mathrm{th}$ percentiles.}
\label{fig:dust_metal_age}
\end{figure}

Dust extinction and stellar metallicities for our clump sample are estimated alongside their stellar ages from the best-fit templates during the SED fitting procedure (Section~\ref{sec:beagle}). Figure~\ref{fig:dust_metal_age} shows the dust extinction ($A_V$) and stellar metallicity ($Z^\star$) as a function of the galactocentric radius (\textit{left panel}) and as evident, there is no significant gradient in either. Furthermore, we directly check for correlation between the stellar age and $A_V$ as well as $Z^\star$, shown in Figure~\ref{fig:dust_metal_age} (\textit{right panel}). While there are some outliers, there is no systematic bias in the best-fit $A_V$ or $Z^\star$ as a function of the best-fit stellar age. Thus, we are confident that our best-fit templates are not preferentially attributing any color gradients in our data to dust or metallicity, instead of the age.

\section{Validating the detectability of the age gradient using the full galaxy disk}
\label{appndx:age_gradient_test}

\begin{figure}[!h]
\centering
\includegraphics[width=0.8\textwidth]{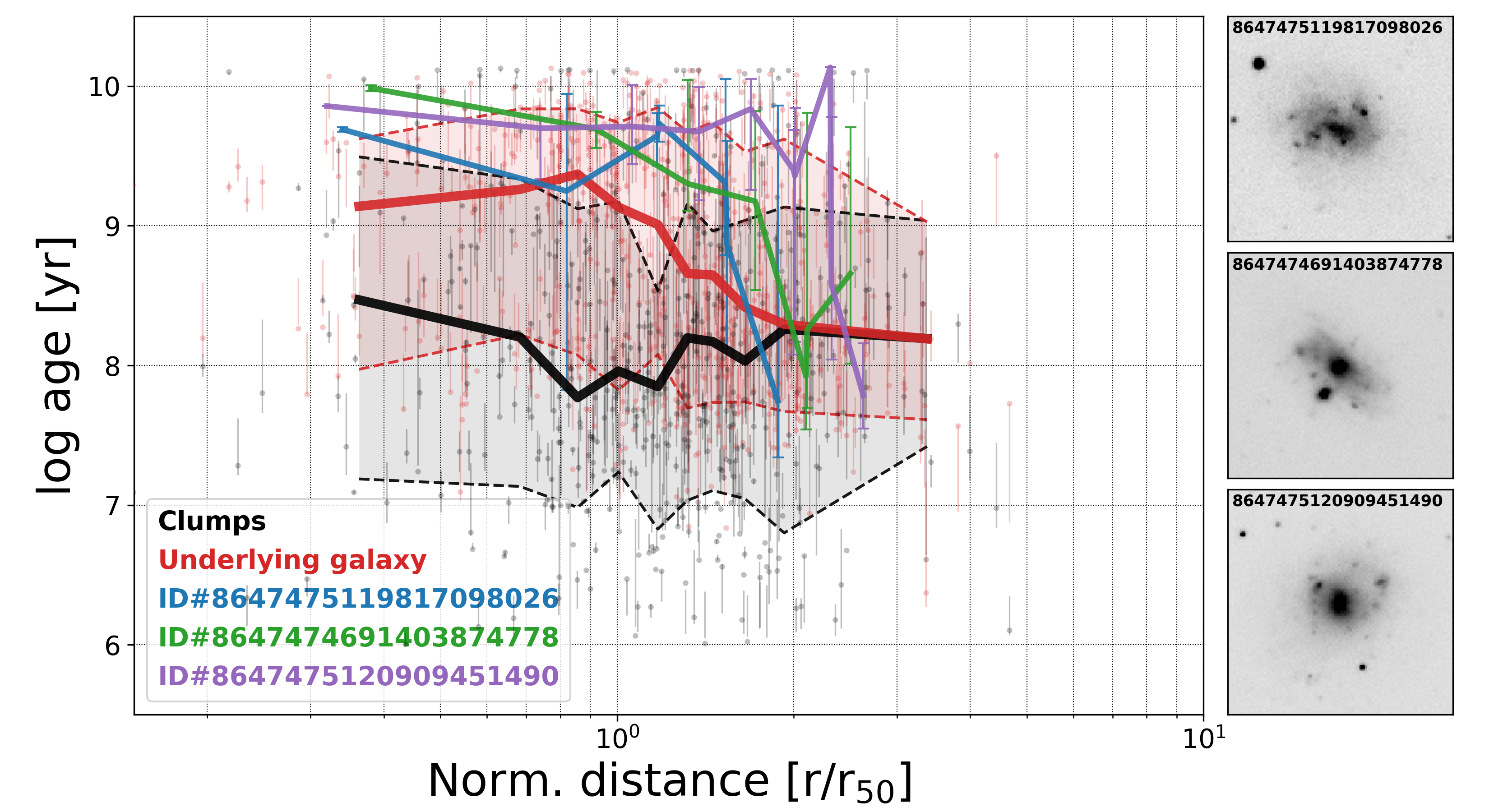}
\caption{Same as the \textit{bottom panel} of Figure~\ref{fig:clump_gradient} with additional curves shown for three example galaxies that have their age gradients computed using the full galaxy disk as opposed to just the disk sampled at the clump locations (\textit{red} curve). The stamps for the three galaxies is shown on the \textit{right}.}
\label{fig:age_gradient_test}
\end{figure}

In Section~\ref{sec:clump_gradient}, we perform a test to validate the ability of our data and methodology to recover any gradients in the clumps' stellar ages, if present. In this section, we extend this validation test to use the full galaxy disk instead of using the galaxy light estimate at the clump location. For this purpose and to keep the test computational tractable, we choose three relatively face-on galaxies (shown in Figure~\ref{fig:age_gradient_test}) and estimate the stellar age gradients using the whole disk. First, we start by masking out all detected clumps in the galaxy and then define apertures by radially dividing the galaxy into concentric annuli where each annulus is further segmented azimuthally to encapsulate any spatial variations within the disk. We measure the photometry for the full disk covered by these aperture segments, thus utilizing almost all of the pixels associated to the galaxy's disk (excluding the clumps). We then use the same SED fitting procedure as for the validity test in Section~\ref{sec:clump_gradient}, which assumes exponentially declining SFHs that are more appropriate for the disk stellar population, to compute the stellar ages for each segment of the galaxy. Finally, the age gradient is then computed by taking the median of the stellar ages for all azimuthal segment in a given annulus.

Figure~\ref{fig:age_gradient_test} shows the age gradients for these 3 example galaxies computed using the full disk. Similar to the test done using the galaxy light at the clump locations, there is a clear gradient visible for each of the three galaxies. The gradient is sharper and occurs at different radii for the individual galaxies. The sharpness of the gradient is expected to be diluted when considering the full galaxy sample as represented by the test using the clump locations, but the resulting gradients are qualitatively similar. The measured age gradients for the individual galaxies are also broadly consistent with those reported in literature \citep[e.g.,][]{gonzalez14,gonzalez15,ruiz17} for local galaxies with similar masses. In conclusion, we have meticulously tested and can confirm that our data and methodology of estimating the stellar parameters is capable of recovering any age gradients, if present.

\bibliographystyle{aasjournal}
\bibliography{mehta_gz_clumps}

\end{document}